\documentclass[11pt]{JHEP3}
\usepackage{latexsym}
\usepackage{graphics}
\usepackage{amsmath}
\usepackage{amsfonts}
\newcommand{\bmat}{\left(\begin{array}}
\newcommand{\emat}{\end{array}\right)}
\newcommand{\beq}{\begin{equation}}
\newcommand{\eeq}{\end{equation}}
\newcommand{\sub}[1]{\phantom{}_{(#1)}\phantom{}}
\def\yzero{\smash{\hbox{$y\kern-4pt\raise1pt\hbox{${}^\circ$}$}}}
\def\p{\partial}
\def\a{\alpha}

\def\g{\gamma}
\def\d{\delta}
\def\Om{\Omega}
\def\om{\omega}

\def\-{\hphantom{-}}

\def\s2{\frac{1}{\sqrt2}}

\def\beq{\begin{equation}}
\def\eeq{\end{equation}}

\def\be{\begin{equation}}
\def\ee{\end{equation}}

\def\beqa{\begin{eqnarray}}
\def\eeqa{\end{eqnarray}}

\def\bea{\begin{eqnarray}}
\def\eea{\end{eqnarray}}

\def\half{\frac{1}{2}}

\def\IF{\relax{\rm I\kern-.18em F}}
\def\II{\relax{\rm I\kern-.18em I}}
\def\IP{\relax{\rm I\kern-.18em P}}

\def\cn{{\cal N}}

\def\Dsl{\,\raise.15ex\hbox{/}\mkern-13.5mu D} %can be subscripted

\def\IC{\bf C}
\def\IZ{\bf Z}

\def\z2z2{$\IC^3/(\IZ_2\times\IZ_2)$}

\def\a{\alpha}

\def\d{\delta}
\def\e{\epsilon}
\def\f{\phi}
\def\g{\gamma}

\def\k{\kappa}
\def\l{\lambda}
\def\m{\mu}
\def\n{\nu}

\def\p{\pi}

\def\r{\rho}
\def\s{\sigma}
\def\t{\tau}
\def\x{\xi}
\def\z{\zeta}
\def\D{\Delta}
\def\F{\Phi}
\def\G{\Gamma}

\def\P{\Pi}

\def\S{\Sigma}

\def\cn{{\cal N}}
\def\co{{\cal O}}

\def\bo{{\raise-.3ex\hbox{\large$\Box$}}}               % D'Alembertian
\def\pa{\partial}                                       % curly d
\def\face{{\raise.2ex\hbox{$\displaystyle \bigodot$}\mskip-2.2mu \llap {$\ddot
        \smile$}}}                                      % happy face
\def\leftrightarrowfill{$\mathsurround=0pt \mathord\leftarrow \mkern-6mu
        \cleaders\hbox{$\mkern-2mu \mathord- \mkern-2mu$}\hfill
        \mkern-6mu \mathord\rightarrow$}       % <--> double differential
\def\dvec#1{\vbox{\ialign{##\crcr
        \leftrightarrowfill\crcr\noalign{\kern-1pt\nointerlineskip}
        $\hfil\displaystyle{#1}\hfil$\crcr}}}           % <--> accent
\def\dt#1{{\buildrel {\hbox{\LARGE .}} \over {#1}}}     % dot-over for sp/sb
\def\beq{\begin{equation}}
\def\eeq{\end{equation}}

\def\beqx{\begin{displaymath}}
\def\eeqx{\end{displaymath}}

\def\beqa{\begin{eqnarray}}
\def\eeqa{\end{eqnarray}}
\def\NO{\nonumber}

\def\>{\rangle} %right angle

\def\<{\langle} %left angle

\title{Correlation Functions in Holographic RG Flows}

\author{Ioannis Papadimitriou and Kostas Skenderis\\ Institute for  Theoretical Physics, 
University of Amsterdam, Valckenierstraat 65, 1018 XE Amsterdam, 
The Netherlands.\\ E-mail: \email{ipapadim@science.uva.nl}, 
\email{skenderi@science.uva.nl}}

\abstract{We discuss the computation of correlation functions in holographic
RG flows. The method utilizes a recently developed Hamiltonian
version of holographic renormalization and
it is more efficient than previous methods. A significant simplification 
concerns
the treatment  of infinities: instead of performing a general
analysis of counterterms, we develop a method where only the 
contribution of counterterms to any given correlators needs
to be computed. For instance, the computation of renormalized
2-point functions requires only an analysis
at the linearized level. We illustrate the method by
discussing flat and AdS-sliced domain walls. 
In particular, we discuss
correlation functions of the Janus solution, a recently discovered
non-supersymmetric but stable AdS-sliced domain wall. }


\preprint{ITFA-2004-23, hep-th/0407071}

\begin{document}

\tableofcontents
\addtocontents{toc}{\protect\setcounter{tocdepth}{2}}

\section{Introduction}
\setcounter{equation}{0}

One of the successes of the gravity/gauge theory correspondence
is the ability to perform detailed computations of QFT correlation
functions by doing a classical gravitational computation.
The holographic prescription put forward in \cite{Gubser:1998bc,Witten:1998qj}
identifies the boundary values of bulk fields with sources of
gauge invariant operators of the boundary theory and the bulk partition
function (which is a functional of the boundary data) with the
generating functional of correlation functions of gauge invariant
operators.
In particular, the supergravity on-shell action is identified
with the generating functional of connected graphs of the (strongly coupled)
boundary QFT at large $N$.

The correlation functions computed using this prescription exhibit
divergences due to the infinite volume of spacetime and need to be
renormalized. A systematic method for doing this was developed in
\cite{dHSS,howtogo,holren} following earlier work in \cite{HS},
see \cite{Skenderis:2002wp} for a review. The method is complete
and can be used to compute renormalized correlation functions in
any holographic RG-flow described by a (locally) asymptotically
AdS (AAdS) spacetime. Some of the steps, however, are rather
tedious which make the method difficult to apply. In this paper we
will present a significantly simpler version of the method.

Let us first recall the computation of 2-point functions
as developed in the early period of the AdS/CFT
correspondence \cite{Gubser:1998bc,fmmr}. We will call this method
the ``old approach''.
To regulate the theory one imposes a cutoff at large radius and solves the
linearized fluctuation equation with Dirichlet boundary condition at
the cutoff. The second variation of
the cutoff on-shell action is then computed in momentum space
yielding an expression
containing singular powers of the cutoff times integer powers of $p$
plus non-singular terms in the cutoff which are non-analytic in $p$.
The 2-point function is defined as the leading non-analytic term.
The polynomial terms in $p$ which are dropped are contact
$\d$-function terms in the position space correlator, which are
scheme dependent in field theory and largely unphysical. The
non-analytic term has an absorptive part in $p$ which correctly
gives the 2-point function for separated points in $x$-space. This method is
quite efficient, but it is not fully satisfactory since it is not correct
in general to simply  drop divergent terms in correlation functions.
The subtractions should be consistent with each other and they
should respect all (non-anomalous) symmetries.

In holographic renormalization one replaces the above computation
by a two step procedure. In the first step renormalization
is performed. This is done via the so-called near-boundary
analysis. This step associates to a given bulk action a
set of universal local boundary counterterms that render
the on-shell action finite on an arbitrary solution of the bulk
equations of motion. Furthermore, to each bulk field
we associate a pair of conjugate variables: the ``source'' and the
exact 1-point function in the presence of sources (or ``vev'').
These can be read off from the asymptotic expansion of the bulk fields.

$n$-point functions can now be obtained by functionally differentiating the
1-point function w.r.t. the sources. The near-boundary analysis
leaves undetermined the vev part of the solution. To obtain
the vev part in terms of the source part we
need an exact (as opposed to asymptotic) solution of the
bulk field equations with arbitrary Dirichlet
data. Such computation is (presently) not possible in all generality
but one can solve the bulk equations perturbatively around a given
solution. In particular, to obtain 2-point functions it is sufficient
to obtain a solution of the linearized field equations, thus
connecting with the method described earlier.
Since all subtractions are made by means of covariant counterterms
mutual consistency is guaranteed. Extra bonus of the method is that
all Ward identities and anomalies are manifest ab initio and
1-point functions (with sources equal to zero) are automatically
computed. Furthermore, the procedure demonstrates that holography
exhibits general field theoretic features such as the fact that
the cancellation of UV divergences does not depend on the IR physics.
The main drawback of the method is that the near-boundary analysis
is somewhat tedious to carry out.

It is conceptually satisfying that the procedure of renormalization
can be carried out in full generality without reference to
a particular solution, i.e. that the counterterms are associated to a
 bulk action, not a particular solution. On the other hand,
however, interesting solutions corresponding to different RG-flows
usually involve different sets of fields and different actions.
To compute correlation functions in different RG-flows one would
thus need to first complete the near-boundary analysis for
each different action.
Furthermore, some counterterms may have vanishing contribution
when we consider a specific solution and may contribute to only
a few correlation functions. Thus, if we are only interested
in the computation of a small set of correlation functions
it would be more efficient
if we had a method that computes only the contribution
of the counterterms to these correlators (rather than first
computing all counterterms and then specializing to the specific
case). The main result of this work is the development
of such a method.

In \cite{PS1}, extending previous work \cite{Kraus:1999di,deBoer:1999xf,Martelli:2002sp} 
(see also \cite{Fukuma:2002sb}), we developed a Hamiltonian method to obtain
renormalized correlation functions of the dual quantum field theory from classical
AdS gravity. The method is based on a Hamiltonian treatment of gravity using the
AdS radial coordinate as the `time' coordinate. More concretely, the
canonical momenta conjugate to the bulk fields evaluated on a regulating
hypersurface are identified with the (regulated) one point functions. These
are functionals of the bulk fields on the same hypersurface and the bulk field
equations become first order functional differential equations for the 
momenta. To obtain covariant counterterms one
 expresses the radial derivative in terms of functional derivatives
w.r.t. the bulk fields. The Dirichlet boundary conditions on the bulk fields then 
imply that the radial derivative is asymptotically identified with the total
dilatation operator of the dual field theory. This is indeed consistent
with the interpretation of the bulk radial coordinate as the energy scale of
the dual field theory. The momenta are then expanded in covariant eigenfunctions
of the dilatation operator, which is the analogue of the asymptotic expansion
of the bulk fields in the standard approach. However, since the expansion is by
construction covariant, all counterterms, including those related to the
conformal anomaly,  are directly computed without the need to invert the 
asymptotic expansions. The method, as the standard method of holographic
renormalization, makes all (anomalous and non-anomalous) Ward identities
manifest and in fact we showed that the  two approaches are completely equivalent.
Nevertheless, the new method leads to a much more efficient algorithm for
the computation of correlation functions from holography as we will demonstrate
in this paper by various examples. The main advantage is that the 
divergences of the holographic correlators can be removed consistently
without the need for a general near boundary analysis as was done in previous
methods.

The paper is organized as follows. In Section 1 we recall some of the
highlights of the method developed in \cite{PS1} to the extent that is necessary 
for the subsequent discussion. In Section 2, we consider Poincar\'{e} domain
wall solutions of the bulk supergravity theory with a single active scalar
field turned on. We discuss how to distinguish between explicit breaking of the 
conformal symmetry by an operator 
deformation and spontaneous breaking by vevs  
by computing  the exact one-point 
functions for these backgrounds. Two point functions are then calculated
and we show explicitly how they can be renormalized without the need for a
general near boundary analysis. Finally, in Section 4 we study the recently 
found Janus solution \cite{BGH}, which is a particular non-supersymmetric but 
stable AdS domain wall \cite{FNSS}. We calculate the vevs of the background
as well as some two-point functions and we show that the Ward identities
associated with the symmetries of the background are satisfied.

\section{Hamiltonian Holographic Renormalization}
\setcounter{equation}{0}

We start by considering the truncated supergravity action
describing gravity in (d+1)-dimensions
coupled to a single scalar field:
\be\label{bulkaction}
S[g,\Phi]=\int_M d^{d+1}x\sqrt{g}\left[-\frac{1}{2\kappa^2}R+
\frac12 g^{\m\n}\partial_\m
\Phi\partial_\n\Phi+V(\Phi)\right]-\frac{1}{2\kappa^2}\int_{\partial
M}d^dx\sqrt{\g}2K.
\ee
where $\partial M$ is the conformal boundary  of the (locally) 
asymptotically AdS spacetime $M$.
By choosing a suitable gauge the bulk metric can be taken to be of the form
\be ds^2=g_{\m\n}dx^\m
dx^\n=dr^2+\g_{ij}(r,x)dx^idx^j,
\ee
where $i=1,\ldots d$. Using
the Gauss-Codazzi relations the bulk equations of motion are
reduced to \cite{PS1}
\beqa
K^2-K_{ij}K^{ij}=R+\k^2
\left[\dot{\Phi}^2-\g^{ij}\partial_i\Phi\partial_j\Phi-
  2V(\Phi)\right],\nonumber \\
\label{greqofmotion}\nabla_iK^i_j-\nabla_jK=\k^2 \dot{\Phi}\partial_j\Phi,\\
\dot{K}^i_j+KK^i_j=R^i_j-\k^2\left[\partial^i\Phi\partial_j\Phi+\frac{2}{d-1}
V(\Phi)\d^i_j\right].\nonumber
\eeqa
Here $K_{ij}=\frac12 \dot{\g}_{ij}$ is the extrinsic curvature and $\dot{K}^i_j$
 stands for $\frac{d}{dr}(\g^{ik}K_{kj})$. Additionally, we have the equation of motion
for the scalar field
\be \ddot{\Phi}+K\dot{\Phi}+\square\Phi-V'(\Phi)=0
\ee
and the on-shell action takes the form \cite{PS1}
\beq \label{on_shell_action}
S_{\rm on-shell}=-\frac{1}{\k^2}\int_{\S_{r}}d^dx\sqrt{\g}(K-\l),
\eeq
where $\l$ satisfies the equation
\be \dot{\l}+K\l+\frac{2\k^2}{d-1}V(\Phi)=0.
\ee

The  Hamiltonian formalism allows one to write
the canonical momenta on a regulating surface $\S_r$, diffeomorphic to the
boundary $\partial M$ at $r\to\infty$, as functional derivatives of the on-shell
action on $\S_r$ w.r.t. the corresponding bulk field. This shows that the canonical
momenta (on $\S_r$) are covariant functionals of the induced fields on $\S_r$, 
which in turn leads to an
expression for the radial derivative in terms of functional derivatives:
\be
\partial_r=\int d^dx 2K_{ij}[\g,\F]\frac{\d}{\d\g_{ij}}+
\int d^dx\dot{\F}[\g,\F]\frac{\d}{\d\F}.
\ee
The asymptotic behavior of the bulk fields then implies that as $r\to\infty$
\be
\partial_r\sim\int d^dx 2\g_{ij}\frac{\d}{\d\g_{ij}}+
\int d^dx(\D-d)\F\frac{\d}{\d\F}\equiv\d_D,
\ee
which is precisely the dilatation operator. This observation is at the center of
our approach.

The next step is to expand the canonical momenta and the on-shell action in
eigenfunctions of the dilatation operator as
\beqa\label{cov_expansions}
K^i_j[\g,\F]=K\sub{0}^i_j+K\sub{2}^i_j+\cdots+K\sub{d}^i_j+\tilde{K}
    \sub{d}^i_j\log{\rm e}^{-2r}+\cdots, \nonumber \\
\l[\g,\F]=\l\sub{0}+\l\sub{2}+\cdots+\l\sub{d}+\tilde{\l}\sub{d}\log{\rm e}^{-2r}+
\cdots,\\
\p[\g,\F]=\dot{\F}[\g,\F]=\sum_{d-\D\leq s< \D} \p\sub{s}+\p\sub{\D}+\tilde{\p}
\sub{\D}\log{\rm e}^{-2r}+\cdots, \NO
\eeqa
where
\beqa\label{transformations}
\d_DK\sub{n}^i_j=-n K\sub{n}^i_j,\,\,n<d,\,\,\,\d_D\tilde{K}\sub{d}^i_j=
    -d \tilde{K}\sub{d}^i_j,\nonumber \\
\d_DK\sub{d}^i_j=-d K\sub{d}^i_j-2\tilde{K}\sub{d}^i_j,\NO \\
\d_D\l\sub{n}=-n \l\sub{n},\,\,n<d,\,\,\,\d_D\tilde{\l}\sub{d}=
    -d \tilde{\l}\sub{d},\nonumber \\
\d_D\l\sub{d}=-d\l\sub{d}-2\tilde{\l}\sub{d}, \nonumber\\
\d_D\p\sub{s}=-s \p\sub{s},\,\,d-\D\leq s<\D,\,\,\,\d_D\tilde{\p}\sub{\D}=
    -\D \tilde{\p}\sub{\D},\nonumber \\
\d_D\p\sub{\D}=-\D\p\sub{\D}-2\tilde{\p}\sub{\D}.
\eeqa
In particular, the coefficients of the logarithmic terms are related to
the conformal anomaly\footnote{
Recall that the gravitational part of the conformal anomaly 
is non-zero only for even dimension $d$ but matter conformal
anomalies \cite{Petkou:1999fv} are generically non-zero for all $d$.
Correspondingly, the logarithmic terms for pure gravity are non-zero only
for even boundary dimension $d$, but they are generically non-zero
even for odd $d$ when additional fields are turned on \cite{dHSS}.}, 
while the terms which transform inhomogeneously under
the dilatation operator are in general non-local and correspond to the
exact one-point functions of the dual operator. More precisely, we have
\be
\langle T_{ij}\rangle_{\rm ren}=-\frac{1}{\k^2}\left(K\sub{d}_{ij}-
K\sub{d}\g_{ij}\right),\;\;\;\;\;\;
\langle\mathcal{O}\rangle_{\rm ren}=\frac{1}{\sqrt{\gamma}}\p\sub{\D}.
\ee
At the same time, the renormalized on-shell action is given by
\be
S_{\rm ren}=-\frac{1}{\k^2}\int_{\S_{r}}d^dx\sqrt{\g}(K\sub{d}-\l\sub{d}).
\eeq
The second equation in (\ref{greqofmotion}), then leads to the Ward identity
\be\label{Ward1}
\nabla_i\langle T^i_j\rangle_{\rm ren}=-\langle\mathcal{O}\rangle_{\rm ren}\partial_j\F,
\ee
whereas the following relation, which was derived in \cite{PS1},
\be
(1+\d_D)K\sub{d}+\k^2(\D-d)\p\sub{\D}\F=(d+\d_D)\l\sub{d}
\ee
gives the (anomalous) Ward identity
\be\label{Ward2}
\langle T^i_i\rangle_{\rm ren}=(\D-d)\F\langle\mathcal{O}\rangle_{\rm ren}+\mathcal{A}.
\ee
Here $\mathcal{A}=-\frac{2}{\k^2}(\tilde{K}\sub{d}-\tilde{\l}\sub{d})$ is the conformal
anomaly defined by
\be
\d_DS_{\rm ren}=-\int_{\S_r}d^dx\sqrt{\g}\mathcal{A}.
\ee
As a consistency check, notice that the above transformation rules
(\ref{transformations}) imply that the conformal anomaly is conformally
invariant as required:
\be
\d_D^2S_{\rm ren}=0.
\ee

A general algorithm for the recursive evaluation of the coefficients in
the covariant expansions was given in \cite{PS1}. It leads to general
expressions for covariant counterterms in arbitrary dimension for pure
gravity, but in general the details depend on the matter that couples
to gravity and so we will skip the details except when necessary to
illustrate the examples we give below. A final comment is due regarding
the renormalization scheme dependence of the correlators we have
computed. One could add extra finite covariant boundary terms in the action
(\ref{bulkaction}) without altering the equations of motion. The corresponding
terms in the covariant expansions would then be shifted accordingly and so
would be the one-point functions. This is all in complete agreement with
our expectations from the field theory point of view.

\section{Poincar\'e Domain Walls}
\setcounter{equation}{0}

In this section we will consider linear fluctuations around Poincar\'e domain
wall solutions of the equations of motion. These take the generic form
\be ds_B^2=dr^2+e^{2A(r)}dx^i dx^i,\;\;\;\;\Phi=\f_B(r).
\ee
Inserting this ansatz into the equations of motion we find that $A(r)$ and
$\f_B(r)$ satisfy
\beqa
\dot{A}^2-\frac{\k^2}{d(d-1)}\left(\dot{\f}_B^2-2V(\f_B)\right)=0,\\
\ddot{A}+d\dot{A}^2+\frac{2\k^2}{d-1}V(\f_B)=0,\\
\ddot{\f}_B+d\dot{A}\dot{\f}_B-V'(\f_B)=0.
\eeqa
Moreover, one has
\be
\dt{\l}_B+d\dot{A}     \l _B+\frac{2\k^2}{d-1}V(\f_B)=0,
\ee
which immediately implies
\be
\l_B=\dot{A}+\x(x) e^{-dA}=\frac1d K_B+\frac{\x(x)}{\sqrt{\g_B}},
\ee
where $\x$ is an arbitrary integration function of the transverse
coordinates. Therefore,
\be
S_{\rm on-shell}^B=-\frac{d-1}{d\k^2}\int_{\S_{r}}d^dx\sqrt{\g_B}K_B+
\frac{1}{\k^2}\int_{\S_{r}}d^dx\x(x).
\ee
The last term corresponds to finite local counterterms.
As expected, there is an ambiguity on the on-shell value of the action 
corresponding to the renormalization scheme dependence of the dual
field theory.

It is well known that the second order flat domain wall equations are solved by
any solution of the first order flow equations 
\beqa\label{floweqs}
\dot{A}=-\frac{\k^2}{d-1}W(\f_B),\NO\\
\dot{\f}_B=W'(\f_B),
\eeqa
provided the potential can be written in the form
\be\label{superpotential}
V(\f_B)=\frac12\left[W'{}^2-\frac{d\k^2}{d-1}W^2\right].
\ee
The motivation for considering theories with such potentials stems from
the fact that they guarantee gravitational stability of the AdS
critical point and of associated domain-wall spacetimes, provided
the AdS critical point is also a critical 
point of $W$ \cite{Townsend,ST,FNSS}. This means,
by the AdS/CFT duality, that the dual theory is unitary. Notice that
in principle it is always possible to write the potential in the form
(\ref{superpotential}) if one views (\ref{superpotential}) as
a differential equation for $W(\f_B)$. The resulting $W$ however may not
have the original AdS spacetime as a critical point. Furthermore, in practice it
is considerably difficult to solve (\ref{superpotential}). 
As a first step in this direction we
observe that (\ref{superpotential}) can be transformed into Abel's equation \cite{PS1}.
In general, the set of solutions of the second order equations 
may include solutions which cannot be obtained from the first order flow equations.
In this section we will restrict attention to solutions which can be derived from
the flow equations. For this class of solutions we have
\be\label{onshellactionbk}
S_{\rm on-shell}^B=\int_{\S_{r}}d^dx\sqrt{\g_B}W(\f_B)+
\frac{1}{\k^2}\int_{\S_{r}}d^dx\x(x).
\ee

\subsection{Deformations vs VEVs}

The domain wall solutions we have described correspond via the AdS/CFT
duality to deformations of the boundary CFT by relevant
operators, in which case the conformal invariance of the boundary theory is
explicitly broken, or to a vacuum expectation value of a scalar operator which
spontaneously breaks the conformal symmetry. Marginal deformations are also
similarly described.

Locally asymptotically AdS metrics satisfy the asymptotic condition $\g_{ij}(r,x)
\sim e^{2r}\g_{ij}\sub{0}(x)$ as $r\to\infty$, which requires $A(r)\sim r$.
Moreover, a scalar field dual to an operator of dimension $\D$ behaves
asymptotically as $\F(r,x)\sim e^{-(d-\D)r}\f\sub{0}(x)$. These asymptotic
conditions, together with the equations of motion, require that the scalar
potential takes the form
\be
V(\F)=-\frac{d(d-1)}{2\k^2}+\frac12 m^2\F^2+\ldots
\ee
where the mass is related to the dimension $\D$ of the dual operator by
$m^2=\D(\D-d)$. Solving the equation of motion for the scalar field with
such a potential leads to a generic solution of the form
\be \label{Fexp}
\F(r,x)=e^{-(d-\D)r}\left[\f\sub{0}(x)+\ldots\right]+
e^{-\D r}\left[\f\sub{2\D-d}(x)+\ldots\right].
\ee
We will consider operators for which $d-\D\geq \D$, or $\D\geq d/2$. Moreover,
we are interested in relevant or marginal operators and so $\D\leq d$. In total
then $d/2\leq\D\leq d$. In this range the first term in the solution for
$\F$ is dominant asymptotically and corresponds to the source, while the
second term is related to the one-point function of the dual operator. In the
special case $\D=d/2$, which saturates the BF bound, the scalar field takes the
form\footnote{Note that
the radial coordinate we use here
is related to the Fefferman-Graham radial coordinate by $\rho=e^{-2 r}$. The factor
$-2r=\log\r$ is chosen to match with  
 the corresponding formulae in \cite{howtogo,holren}. \label{conv}
}
\be\label{BFsaturated}
\F(r,x)=e^{-dr/2}\left[-2r\left(\phi\sub{0}(x)+\ldots\right)+
\tilde{\f}\sub{0}(x)+\ldots\right].
\ee
Again, the first term is the source for the dual operator, while the second
is related to its expectation value.

Depending on the form of the `superpotential' $W(\f_B)$ the domain wall solution
can describe either a deformation of the dual CFT or a phase with spontaneously
broken conformal symmetry as we now explain. A similar analysis can be found in 
\cite{Martelli:2002sp}. Assuming the potential has a critical
point at $\f_B=0$, equation (\ref{superpotential}) together with the flow equations
(\ref{floweqs}) and the requirement that $\f_B=0$ is also a critical point of $W$,
imply that $W$ has an expansion around $\f_B=0$ of two possible forms:
\beqa
W_+(\f_B)=-\frac{d-1}{\k^2}-\frac12(d-\D)\f_B^2+\ldots \\
W_-(\f_B)=-\frac{d-1}{\k^2}-\frac12\D\f_B^2+\ldots.
\eeqa
Which of these cases is realized is purely a property of the background
solution and we need to examine each case separately. Moreover, the extremal
values $d/2$ and $d$ of the scaling dimension $\D$ require special attention.
We will now compute the vev for the stress energy tensor and dual scalar
operator for all cases.

As was shown in \cite{PS1}, the part of the divergent part 
of the on-shell action involving only the scalar field 
is a function $U(\F)$, satisfying the equation
for the `superpotential' $W$ (\ref{superpotential}), {\em and}
having an expansion around $\F=0$
\be
U(\F)=-\frac{d-1}{\k^2}-\frac12(d-\D)\F^2+\ldots
\ee

Let us consider first the case $W_+$ is realized in the background.
In this case we can choose a scheme where the counterterm action 
is $W_+$. To see this notice that 
any two solutions of (\ref{superpotential}) with identical expansions
around $\F=0$ up to order $\F^2$ can only differ at order $\F^{d/(d-\D)}\sim
e^{-dr}$. This is easily proved by looking for the most general power
series\footnote{As usual one must include a logarithmic term at order $d/(d-\D)$
in the general case.}
solution of (\ref{superpotential}) with this particular form up to quadratic
order in $\F$. One finds that all terms are uniquely determined up to order
$d/(d-\D)$ where the recursion relations break down. This is precisely where
an arbitrary integration constant appears and the two solutions could be
potentially different. However, this is irrelevant for the purpose of removing
the divergences of the on-shell action and so we can choose the renormalization
scheme $U(\F)=W_+(\F)$,  and set  the integration function $\x$ to zero. 
This choice of counterterms
corresponds to a supersymmetric renormalization scheme since it ensures
$S_{\rm ren}^B=0$ \cite{howtogo}. 
It follows that the background vevs of both the operator dual
to $\F$ and the stress tensor vanish identically and so this background describes
a deformation of the boundary CFT by a relevant operator.

Let us now consider the case $W_-$ is realized in the background.
In this case we cannot choose $W_-$ as the counterterm since 
it differs from  $U(\F)$ at the quadratic order. In this case, however,
$\f_B\sim e^{-\D r}$ and so the on-shell action evaluated on the background
contains only the volume divergence since, by the hypothesis, $2\D>d$. Hence,
again, setting $\x=0$ corresponds to a supersymmetric renormalization scheme
with $S_{\rm ren}^B=0$. Accordingly, the background expectation value of the
stress tensor vanishes, but not that of the scalar operator. In this case
$W'_-(\f_B)-U'(\f_B)=-(2\D-d)\f_B+\ldots$, and hence,
\be
\langle \mathcal{O}\rangle_{\rm ren}^B=(d-2\D)\f_B,
\ee
which spontaneously breaks the conformal symmetry of the dual CFT. Here we used the
fact that the regularized 1-point function (=canonical momentum) is related to
the superpotential via the first order equation (\ref{floweqs}). The renormalized
1-point function is obtained by subtracting the contribution $U'$ of the counterterm.

It remains to examine the two extremal cases $\D=d/2$ and $\D=d$. When $\D=d/2$
there is no distinction between $W_+$ and $W_-$ as they are equal and
$\f_B\sim e^{-dr/2}$, i.e. it behaves asymptotically as the vev term in
(\ref{BFsaturated}). However, the on-shell action function $U(\F)$ includes
divergences coming from the source term in (\ref{BFsaturated}) and therefore
it cannot be identified with $W(\F)$. It is straightforward to find the
covariant counterterms for this case with the method of \cite{PS1}, but
the singularity structure for the terms involving the scalar field are
not the standard ones, as we now explain. This can be traced back to the fact
that the source term for the scalar contains a logarithm, in contrast to
the generic case. A simple calculation using the asymptotic
form of the solution shows that the canonical momentum
of the scalar field takes the form
\be
\p=\dot{\F}=\left(\frac{1}{r}-\frac{d}{2}\right)\F+\ldots
\ee
Hence, the on-shell action is
\be
S_{\rm on-shell}=S_{\rm on-shell}^{\rm gr}+\int_{\S_r}d^dx\sqrt{\g}
\frac12\left(\frac{1}{r}-\frac{d}{2}\right)\F^2+\ldots
\ee
where $S_{\rm on-shell}^{\rm gr}$ is the on-shell action for pure gravity.
We have therefore shown that
\be
U(\F)=-\frac{d-1}{\k^2}+\frac12\left(\frac{1}{r}-\frac{d}{2}\right)\F^2+\ldots
\ee
which is the divergent part of the on-shell action and must be removed
(of course there is an other part coming from pure gravity). The same
counterterms (for $d=4$) were derived in \cite{howtogo,holren}.
We now see that the scalar operator gets a vev since
$W'(\f_B)-U'(\f_B)=-\frac{1}{r}\f_B  +\ldots$, i.e.
\be
\langle\mathcal{O}\rangle_{\rm ren}^B=2\f_B  .
\ee
This also agrees with the results in \cite{howtogo,holren}. 
Again the difference between $W(\f_B)$ and $U(\f_B)$ is subleading
and  the stress tensor gets no vev since $S_{\rm ren}=0$.

Finally we consider the case $\D=d$, for which
\be
\F(r,x)=\left[\f\sub{0}(x)+\ldots\right]+
e^{-d r}\left[\f\sub{d}(x)+\ldots\right].
\ee
The equations of motion require $V'(\F)=0$ and so the potential is just the
cosmological constant $V(\F)=-\frac{d(d-1)}{2\k^2}$. It follows that the
on-shell function $U(\F)$ is also a constant, i.e. the first term of $W_\pm$.
In this case, however, the general solution to (\ref{superpotential}) can
be easily obtained. There are two distinct solutions (cf. eq. 
(2.10) in \cite{FNSS}),
\be
W_+=-\frac{d-1}{\k^2},\phantom{more  }W_-=-\frac{d-1}{\k^2}\cosh
\left(\sqrt{\frac{d\k^2}{d-1}}(\phi-\phi_o)\right)
\ee
in agreement with our general analysis. Notice that in this case
the supergravity action and hence the second order field equations
are invariant under constant shifts of the scalar field. Such a
constant corresponds to the source term of the solution. One may use 
this symmetry to set $\phi_0$ to zero in $W_-$.
Then, exactly as for the case $d/2<\D<d$, if $W_+$ is realized in the
background, then neither the scalar operator nor the stress tensor
acquire a vev, and if $W_-$ is realized, the scalar operator gets a
vev $-d\f_B$, while the vev of the stress energy tensor vanishes.

So finally we
can summarize all possibilities for flat domain wall backgrounds in
the following table:

\begin{center}
\begin{tabular}{|c|c|c|c|}\hline\hline
$\D$ & $W$ & $\langle\mathcal{O}\rangle_{\rm ren}^B$ &
$\langle T^i_j\rangle_{\rm ren}^B$\\ \hline
$d/2<\D\leq d$ & $+$ & 0 & 0 \\ \hline
 & $-$ & $(d-2\D)\f_B$ & 0 \\ \hline\hline
$d/2$ & $\pm$ & $2\f_B$ & 0 \\
\hline\hline
\end{tabular}
\end{center}

\subsection{Linearized Equations}

Now let us consider fluctuations around the backgrounds we have described
so far. We will only keep terms up to linear order in fluctuations, which
suffices for the calculation of the two point functions. The metric
fluctuations take the form
\be
\g_{ij}=\g_{ij}^B(r)+h_{ij}(r,x)=e^{2A(r)}\d_{ij}+h_{ij}(r,x),
\ee
and the scalar field is
\be
\F=\f_B(r)+\f(r,x).
\ee
The extrinsic curvature then becomes
\be
K^i_j=\dot{A}\d^i_j+\frac12\dot{S}^i_j,
\ee
where $S^i_j\equiv \g^{ik}_Bh_{kj}$. $S^i_j$ can be decomposed into irreducible
components as
\be
S^i_j=e^i_j+\partial^i\e_j+\partial_j\e^i+\frac{d}{d-1}
\left(\frac{1}{d}\d^i_j-\frac{\partial^i\partial_j}{\square_B}\right)f+
\frac{\partial^i\partial_j}{\square_B}S,
\ee
where $\partial_i e^i_j=e^i_i=\partial_i\e^i=0$ and indices are raised
with the inverse background metric $e^{-2A}\d^{ij}$. Conversely, each of
the irreducible components can be expressed uniquely in terms of $S^i_j$ as
\be
e^i_j=\P^i\phantom{}_k\phantom{}^l\phantom{}_j S^k_l,\,\,\,\,
\e_i=\p^l_i\frac{\partial_k}{\square_B}S^k_l,\,\,\,\,f=\p^l_k S^k_l,\,\,\,\,
S=\d^l_k S^k_l,
\ee
where we have introduced the projection operators
\be
\P^i\phantom{}_k\phantom{}^l\phantom{}_j=\frac12\left(\p^i_k\p^l_j+
\p^{il}\p_{kj}-\frac{2}{d-1}\p^i_j\p^l_k\right),
\ee
and
\be
\p^i_j=\d^i_j-\frac{\partial^i\partial_j}{\square_B}.
\ee
With this nomenclature we can now go on and derive the equations of motion
for the linear fluctuations. The result is\footnote{Note
$\square_B=e^{-2A}\square=e^{-2A}\d^{ij}\partial_i\partial_j$.}
\beqa
\left(\partial_r^2+d\dot{A}\partial_r+e^{-2A}\square\right) e^i_j=0,\\
\left(\partial_r^2+[d\dot{A}+2W\partial^2_\f\log W]\partial_r+
e^{-2A}\square\right)\om=0,\\
\dot{f}=-2\k^2\dot{\f}_B\f,\\
\dot{S}=\frac{1}{(d-1)\dot{A}}\left[-e^{-2A}\square f+
2\k^2\left(\dot{\f}_B\dot{\f}-V'(\f_B)\f\right)\right]
\eeqa
where
\be
\om\equiv \frac{W}{W'}\f+\frac{1}{2\k^2}f
\ee
and we have used the diffeomorphism invariance in the transverse space
to set $\e_i\equiv 0$. The last two equations give immediately the
momenta dual to $f$ and $S$ and hence the corresponding one-point
functions with linear sources. Moreover, since the canonical momenta
are functionals of the bulk fields \cite{PS1}, to linear order in the
fluctuations we must have
\be
\dot{e}^i_j=E(A,\f_B)e^i_j,\,\,\,\,\,\dot{\om}=\Om(A,\f_B)\om.
\ee
The first two equations then become first order equations for $E$ and $\Om$:
\beqa\label{firstorderequations}
\dot{E}+E^2+d\dot{A}E-e^{-2A}p^2=0,\NO \\
\dot{\Om}+\Om^2+[d\dot{A}+2W\partial^2_\f\log W]\Om-e^{-2A}p^2=0,
\eeqa
where we have performed a Fourier transform in the transverse space. Given
the solutions for $E$ and $\Om$ we can immediately write down all
momenta, namely
\beqa
\dot{e}^i_j=Ee^i_j,\\
\dot{f}=-2\k^2\dot{\f}_B\f,\\
\dot{\f}=(W''+\Om)\f+\frac{1}{2\k^2}\frac{W'}{W}\Om f,\\
\dot{S}=-\frac{1}{\k^2}\left[\left(\frac{W'}{W}\right)^2\Om-
\frac{e^{-2A}}{W}\square\right]f-2\frac{W'}{W}
\left(\Om+\frac{d\k^2}{d-1}W\right)\f.
\eeqa
To completely determine the one-point functions with linear sources
we first need to obtain exact solutions for $E$ and $\Om$ and secondly,
to determine the covariant counterterms for the momenta, {\em but} only
to linear order in the fluctuations. Since $e^{-2A}$ and $W(\f_B)$ are
already covariant functions of the background fields, it suffices to
find covariant expansions for $E$ and $\Om$ in the background fields.
These can be organized according to the dilatation operator for the
background
\be
\d_D=\partial_A+(\D-d)\f_B\partial_{\f_B}.
\ee
More generally,
the radial derivative is expanded in functional derivatives w.r.t.
the background fields as
\be
\partial_r=\dot{A}\partial_A+\dot{\f_B}\partial_{\f_B}=
-\frac{\k^2}{d-1}W(\f_B)\partial_A+W'(\f_B)\partial_{\f_B}\sim \d_D+\ldots
\ee
Inserting the following expansions\footnote{These
expansions are strictly correct for $d/2<\D\leq d$, but we will deal
with the special case $\D=d/2$ in the examples below.}
 for $E$ and $\Om$ in the first order
equations (\ref{firstorderequations}),
\beqa\label{expansions}
E=E\sub{1}+\cdots+\tilde{E}\sub{d}\log(e^{-2r})+E\sub{d}+\cdots,\NO \\
\Om=\Om\sub{0}+\cdots+\tilde{\Om}\sub{2\D-d}\log(e^{-2r})+
\Om\sub{2\D-d}+\cdots,
\eeqa
one determines all covariant counterterms which render all momenta
finite to linear order in the sources. This procedure is substantially
simpler than the general holographic renormalization required to
determine the full non-linear counterterms and is a significant
improvement over  previous methods.

A final simplification can be made for the case of backgrounds
corresponding to deformations of the dual CFT. As we saw in the
previous section, the `superpotential' $W$ of the background can be
included in the counterterm action, corresponding to a supersymmetric
renormalization scheme. After this counterterm is added to the
on-shell action (we still have to determine the counterterms for
$E$ and $\Om$), the momenta take the simpler form
\beqa
\dot{e}^i_j=Ee^i_j,\\
\dot{f}=0,\\
\dot{\f}=\Om\f+\frac{1}{2\k^2}\frac{W'}{W}\Om f,\\
\dot{S}=-\frac{1}{\k^2}\left[\left(\frac{W'}{W}\right)^2\Om-
\frac{e^{-2A}}{W}\square\right]f-2\frac{W'}{W}\Om\f.
\eeqa

\subsection{Examples}

We will treat the two examples that have been the main 
testing ground for holographic computation of correlation
functions, namely the GPPZ flow \cite{GPPZ} and the Coulomb branch flow \cite{CB,Brandhuber:1999}.
The computation of certain two-point functions for the CB flow was first  discussed
in \cite{CB} and for the GPPZ flow in \cite{Anselmi:2000fu}. 
Two-point functions for both flows were systematically studied in 
\cite{mueck,howtogo}, 
see also \cite{Notes,theisen,Anatomy} for earlier work.
Since the results are known, the emphasis
here will be in method rather than the correlators themselves.

{\bf GPPZ Flow}

The GPPZ flow describes a deformation by a supersymmeric mass
term of $\mathcal{N}=4$ SYM. The bulk theory is that of a
scalar field dual to an operator of dimension
$\D=3$ coupled to gravity in five dimensions. The background
`superpotential' is
\be
W(\f_B)=-\frac{3}{2\k^2}\left[1+\cosh\left(\sqrt{\frac{2}{3}}
\k\f_B\right)\right]=-\frac{3}{\k^2}-\frac12\f_B^2+\cdots
\ee
which is of the form $W_+$ and corresponds to a deformation
of the boundary CFT by a relevant operator. The background solution
takes the form
\be
\f_B=\frac{1}{\k}\sqrt{\frac32}\log\left(\frac{1+\sqrt{1-u}}{1-
\sqrt{1-u}}\right),\,\,\,\,\,e^{2A}=\frac{u}{1-u},\,\,\,\,\,1-u=e^{-2r}.
\ee
It is also useful to note the relations
\be
W=-\frac{3}{\k^2}\frac{1}{u},\,\,\,\,W'=-\frac{\sqrt{6}}{\k}
\frac{\sqrt{1-u}}{u},\,\,\,\,W''=\frac{2\k^2}{3}W+1.
\ee
Changing variable from $r$ to $u$ in (\ref{firstorderequations}) we
obtain
\beqa
2(1-u)E'(u)+E^2+\frac4u E-\frac{1-u}{u}p^2=0,\\
2(1-u)\Om'(u)+\Om^2+\left(\frac4u-2\right)\Om-\frac{1-u}{u}p^2=0.
\eeqa
The solutions which are regular at $u=0$ are
\be
E(u)=\frac14p^2(1-u)\frac{F\left(1-\frac{ip}{2},1+
\frac{ip}{2};3;u\right)}{F\left(-\frac{ip}{2},
\frac{ip}{2};2;u\right)}
\ee
and
\be
\Om(u)=\frac14p^2(1-u)\frac{F\left(\frac{3-\a}{2},
\frac{3+\a}{2};3;u\right)}{F\left(\frac{1-\a}{2},
\frac{1+\a}{2};2;u\right)}
\ee
where $\a=\sqrt{1-p^2}$.

Next we need to find covariant counterterms for $E$ and $\Om$.
Inserting
\be
\partial_r=\d_D+\frac{\k^2}{6}\f_B^2\left(\partial_A-\frac23
\f_B\partial_{\f_B}\right)+\cdots
\ee
and the expansions (\ref{expansions}) in
(\ref{firstorderequations}) one very easily determines
\beqa
E=\frac{p^2}{2}e^{-2A}+\frac{p^2}{4}e^{-2A}
\left(\frac{p^2}{2}e^{-2A}+\frac{\k^2}{3}\f_B^2\right)\log e^{-2r}
+E\sub{4}+\cdots,\\
\Om=-\frac{p^2}{2}e^{-2A}\log e^{-2r}+\Om\sub{2}+\cdots.
\eeqa
Expanding the exact solution in $1-u$ and removing  the
covariant terms we have just determined allows for the
evaluation of $E\sub{4}$ and $\Om\sub{2}$, which are
precisely the terms required to calculate the renormalized
one-point functions. Putting everything together we find the
following two-point functions:
\bea
&&\<\co(p) \co(-p)\> = -\frac12 p^2 \bar{J},\qquad
\<T^i_i(p) \co(-p)\> = \frac{\sqrt{6}}{2 \k} p^2  \bar{J}, \NO \\
&&\<T^i_i(p) T^i_i(-p)\> = -\frac{3}{\k^2} p^2 (\bar{J} +1), 
\qquad  p^jp_i\<T^i_j\>=0,\nonumber \\ 
&&\<T_{ij}(p) T_{kl}(-p)\>_{TT} = \frac{2}{\k^2} \P_{ijkl} %{N^2\over 2 \pi^2}
[\frac{1}{16} p^2 (p^2 + {4}) \bar{K} + \frac{p^2}{8}] \NO 
\eea
where
\bea
\bar{J}&=&2 \psi(1) - \psi(\frac{3}{2} +\half \sqrt{1-p^2})
-\psi(\frac{3}{2} - \half \sqrt{1-p^2}) \NO \\
\bar{K}&=&\psi(1) + \psi(3) -\psi(2+\frac{ip}{2})-
\psi(2-\frac{ip}{2}) \nonumber
\eea

{\bf Coulomb Branch Flow}

The Coulomb branch flow is a solution of five dimensional AdS gravity
coupled to a scalar field of mass $m^2=-4$, which therefore
saturates the BF bound. The solution describes the case where an
operator of dimension 2 gets a vev. The superpotential is
\be
W(\f_B)=-\frac{2}{\k^2}\left[e^{-\k\f_B/\sqrt{3}}+\frac12 e^{2\k\f_B/\sqrt{3}}
\right].
\ee 
The solution can be parametrized by $v\equiv e^{\sqrt{3}\k\f_B}$ as
\be
\dot{v}=2v^{2/3}(1-v),\,\,\,\,e^{-2A}=v^{-2/3}(1-v),\,\,\,W=-\frac{1}{\k^2}
v^{-1/3}(v+2).
\ee
The boundary is located at $v=1$. In terms of $v$ the first order equations
(\ref{firstorderequations}) become
\be
2(1-v)E'(v)+v^{-2/3}E^2+\frac43\left(1+\frac{2}{v}\right)E-p^2v^{-4/3}(1-v)=0,
\ee
and
\be
2(1-v)\Om'(v)+v^{-2/3}\Om^2+\left[\frac43\left(1+\frac{2}{v}\right)-
\frac{12}{v+2}\right]\Om-p^2v^{-4/3}(1-v)=0.
\ee
The solution for $E$ which is regular at $v=0$ is
\be
E(v)=2a(1-v)v^{-1/3}\left[1+\frac{av}{2(a+1)}\frac{F(a+1,a+1;2a+3;v)}
{F(a,a;2a+2;v)}\right],
\ee
where $a=-\frac12+\frac12\sqrt{1+p^2}$. We will not give here
explicitly the exact solution for $\Om$ since it is rather complicated.
To obtain such a closed form solution one must transform the above equation for 
$\Om$ into a soluble form and then obtain $\Om$ implicitly through
the solution of the transformed equation. After obtaining covariant counterterms 
for $E$ and $\Om$ by the method we described above, we can write these in the 
desired form, namely
\be
E=\frac{p^2}{2}e^{-2A}+\frac{p^4}{8}e^{-4A}\log e^{-2r}+\frac{p^2}{2}
e^{-4A}\left[-\frac13+\frac{p^2}{2}\left(\psi(a+1)-\psi(1)\right)\right]+\cdots
\ee
\be
\Om=\frac{1}{r}+\frac{1}{r^2}\left(-\frac{4}{3p^2}+\psi(a+1)-\psi(1)\right)+\cdots.
\ee  
Inserting these expansions into the expressions for the momenta, after
taking into account the effect of the counterterm
\be
U(\F)=-\frac{3}{\k^2}+\frac12\left(\frac1r-\frac d2\right)\F^2=
W(\F)+\frac{1}{2r}\F^2+\cdots,
\ee
we obtain the two-point functions
 \bea
&&\<\co(p) \co(-p)\> = %{N^2 \over 2 \p^2}
\left(4 \psi(1) - 4 \psi(1+a) + \frac{16}{3 p^2}\right)
\qquad
\<T^i_i(p) \co(-p)\> = -\frac{4}{\sqrt{3}\k} %{N^2 \over 2 \p^2}
= 2 \<\co\>_B \NO \\
&&p^i p^j \<T_{ij}(p) \co(-p)\> = 
-\frac{2 }{\sqrt{3}\k} p^2
%{N^2 \over 2 \p^2} 
=  \< \co \>_B p^2 \qquad \<T^i_i(p) T^j_j(-p)\> = 0
\nonumber \\
&&\<T_{ij}(p) T_{kl}(-p)\>_{TT}= \P_{ijkl} %{N^2\over 2 \pi^2}
\frac{p^2}{2\k^2}\left[\frac13-\frac{p^2}{2}\left( \psi(a+1)-\psi(1) \right)
\right] \nonumber
\eea

\section{AdS-sliced Domain Walls}
\setcounter{equation}{0}

AdS-sliced domain walls have also been studied in the literature 
\cite{Lust,Sabra,Behrndt,BGH,FNSS}.
In this case the background is of the form
\be
ds_B^2=dr^2+e^{2A(r)}g_{ij}(x)dx^i dx^i,\;\;\;\; \Phi=\f_B(r),
\ee
where $g_{ij}(x)$ is the metric of Euclidean $AdS_d$ with radius $l$
and  we have set the radius of the bulk $AdS_{d+1}$ equal to 1\footnote{In 
\cite{FNSS} a different convention was used:
the radius of bulk $AdS_{d+1}$ and of the $AdS_{d}$-slice were set
equal to each other.}.
Inserting this ansatz into the bulk equations of motion leads to
the following equations for  $A(r)$ and $\f_B(r)$
\beqa
\dot{A}^2-\frac{\k^2}{d(d-1)}\left(\dot{\f}_B^2-2V(\f_B)\right)
+\frac{1}{l^2}e^{-2A}=0,\\ 
\ddot{A}+d\dot{A}^2+\frac{2\k^2}{d-1}V(\f_B)+\frac{d-1}{l^2}e^{-2A}=0,\\
\ddot{\f}_B+d\dot{A}\dot{\f}_B-V'(\f_B)=0.
\eeqa
Note that as $l \to\infty$ these reduce to the equations for
flat domain walls, as they should. From now on we set $l^2=1.$

\subsection{Janus solution}

A particularly interesting  AdS-sliced domain wall solution
is the dilaton domain wall solution of type IIB supergravity of \cite{BGH}\footnote{
Additional dilatonic deformations have been presented in \cite{Bak}.}. 
This is a non-supersymmetric
regular solution. When reduced to five dimensions,
it solves the field equations  of AdS gravity coupled to a massless scalar with a 
constant potential. Similar solutions exist in all
dimensions \cite{FNSS}. These solutions are of particular interest because 
they 
enjoy non-perturbative stability for a broad class of deformations \cite{FNSS}.
This strongly suggests that they should have a well-defined 
QFT dual. 

Setting $V=-\frac{d(d-1)}{2\k^2}$, the equation for the scalar field
can be trivially integrated to give
\be
\dot{\f}_B=c e^{-dA},
\ee
where $c$ is an arbitrary constant of integration. The remaining equations
imply,
\be \label{Aeqn}
\dot{A}^2=1 - e^{-2 A} + b e^{-2 d A}
\ee
where $b=\frac{c^2\k^2}{d(d-1)}$.  The geometry is non-singular 
provided the parameter $b$ is within the range,
\be \label{brange}
0\leq b<b_0 \equiv \frac{1}{d}\left( \frac{d-1}{d} \right)^{d-1}.
\ee
One can obtain an implicit solution of (\ref{Aeqn}) as
\be\label{radialcoordinate}
r = \int_{A_0}^A \frac{d A}{\sqrt{1 - e^{-2 A} + b e^{- 2 d A}}}
\ee
where $A_0$  is the smallest zero of $P(u)\equiv b u^d - u +1$, where
  $u \equiv e^{-2A}$. This defines half of the geometry,
i.e. the region with  $0 \leq r < \infty$. The other half is obtained
by extending $A(r)$ to negative values of $r$ as an even function,
$A(-r)=A(r)$. 

We can obtain an explicit expression for the bulk metric by 
changing variables from $r$ to $u$. Using
\be
\dot{A}^2=1-u+bu^d,
\ee
we obtain\footnote{An equivalent form of this metric with  
$A$ instead of $u$ as a variable was found by C. N\'{u}\~{n}ez 
(unpublished notes, July 2003).}
\be
\label{exactbackgroundmetric}
ds_B^2=\frac{du^2}{4u^2(1-u+bu^d)}+\frac{1}{u}g_{ij}(x)dx^i dx^i.
\ee
Note that if $b=0$ this is precisely the metric for $AdS_{d+1}$
in the $AdS_d$-slicing parameterization. The range of the $u$-coordinate
depends on the value of the parameter $b$, namely 
 $0\leq u\leq u_o$, where $u_o\geq 1$ with
equality iff $b=0$. We give the explicit form of $u_o$ as a
function of $b$ in appendix \ref{Jage}. In this parameterization the two halves of 
the space,
i.e. $r>0$ and $r<0$, are not distinguished since $u$ is an even function of 
$r$. In particular, the regions at $r\to\pm\infty$ are mapped to $u=0$. 

We discuss in appendix \ref{Jage} the conformal compactification 
of the solution. The conformal boundary consists of two half-spheres
with angular excess joined along their equator \cite{BGH,FNSS}.
We will refer to the joining equator as ``corner''.
In order to calculate correlation functions of the dual field theory
we need to write the Janus metric in the Fefferman-Graham (FG) form. 
Provided the boundary metric is smooth, this is always possible in a 
neighborhood of the boundary but the FG radial 
coordinate may in general not be valid far away from the boundary. 
In the present case, the boundary metric is smooth except for the 
presence of corners. We therefore except to be able to find a FG coordinates
that are well defined in the neighborhood of the boundary 
except perhaps at the corner.

In appendix \ref{appendix_FG} we construct the FG metric to all orders in $b$
for the Janus geometry and determine the range of validity of the 
radial coordinate. We find that the FG coordinates are well defined 
everywhere in a neighborhood of the boundary except on the corner
where the two half-spheres of the boundary meet. In particular, the FG
metric takes the form 
\be
ds_B^2=\frac{1}{z_o^2}\left[dz_o^2+(1+bc_3(x)+\mathcal{O}(b^2))dz_d^2+
(1+bc_4(x)+\mathcal{O}(b^2))dz_a^2\right], 
\ee
where $x\equiv z_d/z_o$ and $z_a$, $a=1,\ldots,d-1$ are the standard
transverse coordinates in the upper half plane parameterization of the 
$AdS_d$ slice. The location of the corner is at $z_d=0$.
The functions $c_3(x)$ and $c_4(x)$ as well as the form of 
the FG metric to all orders in $b$ are given in appendix \ref{appendix_FG}.
As discussed there this coordinate system covers the region
$|x|>x_o=b/\sqrt{2}+\mathcal{O}(b^2)$, so $z_d=0$ only when $z_o=0$.
In other words, this coordinate system does not cover a (radially 
extended) neighborhood of $z_d=0$. 
 
In this coordinate system the background scalar takes the form
\be
\f_B(x)=\f_o+c c_5(x)+\mathcal{O}(c^3),
\ee
where again $c_5(x)$ is given in the appendix. It is significant to point out
here that on the boundary, i.e. $z_o=0$, the value of the scalar field is
a step function in $z_d$, namely
\be 
\f_B(z_d)=\f_o + {\rm sgn}(z_d)c,
\ee
which implies that the coupling of the dual operator is different on the 
two sides of the corner, or `wall', at $z_d=0$.  
These results are sufficient for calculating correlation functions, which we 
do in the next section. 

\subsection{VEVs}

Now that we have determined the appropriate FG coordinate system 
we can carry out the algorithm developed in \cite{PS1} and evaluate
the vevs of the stress tensor and the scalar operator dual to the
dilaton, as well as, all two point functions using perturbation 
theory in $c$. The first step is to define the radial coordinate\footnote{This 
radial coordinate is different from  the original radial coordinate in 
(\ref{radialcoordinate}) but we hope this causes no confusion.}
$r=-\log z_o$ which is used as the `time' coordinate in the Hamiltonian
formalism of \cite{PS1}. Due to the fact that the background 
depends also on the transverse space coordinates, a full asymptotic
analysis is required to determine the covariant counterterms. We will
not give these here but they are easily determined following the procedure
in \cite{PS1}. Evaluating these counterterms on the background using
the following expressions for the non-vanishing components of the
Christoffel symbol and Ricci tensor:
\bea
\G^d_{dd}=\frac{b}{2}e^rc_3'(x)+\mathcal{O}(b^2),\,\,\,\phantom{more}
\G^a_{bd}=\frac{b}{2}e^rc_4'(x)\d^a_b+\mathcal{O}(b^2),
\phantom{more space}\\ \NO
R_{dd}=-\frac{(d-1)b}{2}e^{2r}c_4''(x)+\mathcal{O}(b^2),\,\,\,\,
R_{ad}=R_{ab}=\mathcal{O}(b^2),\,\,\,\,R=-\frac{(d-1)b}{2}c_4''(x),
+\mathcal{O}(b^2)
\eea 
and adding them to the canonical momenta obtained directly by differentiating
the background fields w.r.t. $r$ one obtains the following expressions
for the vevs of the scalar operator and the stress tensor:
\be \label{vev}
\langle \mathcal{O}\rangle_B=c
\frac{z_d}{|z_d|^{d+1}}\,,\phantom{more
space here}\langle T^i_j\rangle_B=0.
\ee
Although the calculation has been
done to leading order in $c$, it is not difficult to show that these results
are in fact exact. The reason is that the coordinate transformation
(\ref{ucoordinate}) ensures that for every power of $b$ there is a factor of
$z_o^{2(d-1)}$ which means that higher order in $b$ terms are subleading and
do not survive when the regulator is removed. This can also be seen from the
exact expressions for the Fefferman-Graham  metric and scalar background 
given in appendix \ref{appendix_FG}, which can be used to obtain the exact 
canonical momenta. Namely, 
\be
K_{Bd}^d=\left(1+\frac{bu^d}{1-u}\right)^{1/2},\phantom{more space}
K_{Bb}^a=\left(u+\sqrt{(1-u)(1-u+bu^d)}\right)\d^a_b
\ee
and
\be
\dot{\phi}_B(x)={\rm sgn}(x)cu^{d/2}\sqrt{1-u}.
\ee
One immediately sees that the vevs given above are in fact
exact, as claimed.

The form of the vacuum expectation values is the one required by the symmetries
of the problem. As shown in \cite{McAvity&Osborn}, the 1-point functions
for a conformal field theory on a flat space with a boundary at $z_d=0$ (which 
breaks the conformal group from $O(1,d+1)$ to $O(1,d)$) are precisely of the form
(\ref{vev}). In the present case we consider the theory on both sides of the wall
 $z_d=0$. The McAvity-Osborn result applies separately to the two regions, 
$z_d>0$ and $z_d<0$, and it gives
\be\label{vevsign} 
\langle \mathcal{O}\rangle_B=
\frac{c_1}{|z_d|^{d}},\,\,\,z_d>0,\phantom{more
space } \langle \mathcal{O}\rangle_B=
\frac{c_2}{|z_d|^{d}},\,\,\,z_d<0.
\ee 
In the present case $c_1=-c_2=c$. 

These considerations suggest \cite{BGH} that the dual field theory for $d=4$ 
is $\cn=4$ SYM 
possibly coupled to non-supersymmetric conformal matter localized at $z_d=0$
and with $g_{YM}$ being different on the two sides of the wall
(similar suggestions can be formulated in all dimensions). This is consistent with 
the symmetries of the model: the presence of the defect breaks the 
symmetries to $O(1,4)$ (i.e. the (Euclidean) conformal group in three dimensions).
It would be interesting to investigate whether there is a classical solution 
of $\cn=4$ SYM coupled to such defect that can reproduce (\ref{vev}), but 
we will not pursue this here.\footnote{{\bf Note added:} A precise proposal for 
the dual theory 
was recently made in \cite{Freedman}: they consider $\cn=4$ SYM theory 
on two half-spaces separated by a planar interface that contains no matter
and with a different coupling constant coupled to specific operator
closely related to the $\cn=4$ Lagrangian density. The field theory computations
in \cite{Freedman} exactly agree (to the extend that they can be compared)
with the holographic computations described in this and next 
subsection.}

\subsection{Two-point functions}

Since the leading correction to the $AdS_{d+1}$ metric is order $c^2$, 
while the leading corrections to the (off diagonal) two-point functions 
are order $c$, we can take the background to be exactly $AdS$ and
consider linear fluctuations driven by a source $\tilde{T}^i_j$ which
is of order $c$. Decomposing the metric fluctuations as was done for
flat domain walls above we derive the following equations for the irreducible
components:
\bea
-\square_g e^i_j=2\k^2\P^i\phantom{}_k\phantom{}^l\phantom{}_j\tilde{T}^k_l,\\
\dot{\e}_j=2\k^2\frac{\p^k_j}{\square}\tilde{T}_{kd+1},\\
\dot{f}=-2\k^2\frac{\partial^k}{\square}\tilde{T}_{kd+1},\\
\dot{S}=\frac{1}{d-1}\left[2\k^2\tilde{T}_{d+1d+1}-e^{-2r}\square f\right],\\
-\square_g\f=\frac{1}{2}\dot{\phi}_B\dot{S}-e^{-2r}\left(S^i_j\partial_i
\partial^j\phi_B+\partial_iS^i_j\partial^j\phi_B-\frac12\partial_jS
\partial^j\phi_B\right).\label{scalarequation}
\eea  
Only the first and the last equations need further analysis as the rest
give immediately the momenta as functions of the linear sources. The
responses for both the transverse traceless metric fluctuation and the scalar 
field fluctuation can be obtained using the massless scalar bulk-to-bulk
propagator
\be
G(\x)=\frac{c_d}{2^dd }\x^d
F\left(d,\frac{d+1}{2};\frac d2+1;\x^2\right)
\ee
which satisfies
\be
-\square_g G(\x)=\d(z,w)=\frac{1}{\sqrt{g}}\d(z-w).
\ee
Here $c_d=\G(d)/(\G(d/2)\p^{d/2})$ and 
$\x=2z_ow_o/(z_o^2+w_o^2+(\vec{z}-\vec{w})^2)$. As $z_o\to 0$
\be
G(\x)\sim \frac{z_o^d}{d}K_d(w,\vec{z})
\ee
where
\be
K_d(w,\vec{z})=c_d\left(\frac{w_o}{w_o^2+(\vec{w}-\vec{z})^2}\right)^d
\ee
is the well known bulk-to-boundary propagator. To complete the
calculation then we need the source $\tilde{T}_{\m\n}$ which is
\bea
\tilde{T}_{\m\n} & =& \partial_\m\Phi\partial_\n\Phi-\frac12 g_{\m\n}g^{\r\s}
\partial_\r\Phi\partial_\s\Phi \NO \\
& = & \partial_\m\phi_B\partial_\n\phi+\partial_\n\phi_B\partial_\m\phi-
g_{\m\n}^{AdS}\dot{\phi}_B\left(\dot{\phi}+\frac{e^{-2r}}{z_d}
\partial_{z_d}\phi\right)+\mathcal{O}(b).
\eea
Here we have used the fact that the background scalar is a function of
$x=z_d/z_o$ which implies $z_d\partial_{z_d}\phi_B=\dot{\phi}_B$.
With a little more algebra the source can be cast in the form
\bea
\tilde{T}_{d+1d+1}=\dot{\f}_B\left(\dot{\f}-\frac{e^{-2r}}{z_d}\partial_{z_d}\f
\right)+\mathcal{O}(b),\NO \\ \NO\\
\tilde{T}_{jd+1}=\dot{\f}_B\left(\partial_j\f-\frac{1}{z_d}\d_{jd}\dot{\f}
\right)+\mathcal{O}(b),\NO \\\NO \\
\tilde{T}_{ij}=\dot{\f}_B\left[\frac{1}{z_d}\left(\d_{id}\partial_j\f+
\d_{jd}\partial_i\f-\d_{ij}\partial_{z_d}\f\right)-e^{-2r}\dot{\f}\d_{ij}
\right]+\mathcal{O}(b).
\eea
Using these sources and the above bulk-to-bulk propagator we can now
evaluate the canonical momenta which give the one-point functions with
linear sources. It is not difficult to show that no counterterms contribute 
to the order $c$ terms of the momenta. It turns out to be easier to obtain 
the two-point functions from the canonical momentum of the graviton by
differentiating w.r.t. the scalar source rather than from the scalar momentum 
and so we only consider the graviton momentum here.
Of course both calculations should give identical results and we show this
explicitly in appendix \ref{appendix_2ptfns}, where we calculate the two-point 
functions from the scalar momentum. 

To obtain the the canonical momentum of the transverse traceless component
of the graviton we note that the inhomogeneous solution to its equation of
motion is
\be
e^i_j=2\k^2\int d^{d+1}w\sqrt{g(w)}G(\x)\P^i\phantom{}_k\phantom{}^l
\phantom{}_j\tilde{T}^k_l(w)
\ee
so that asymptotically
\be
\dot{e}^i_j\sim -2\k^2e^{-dr}\int d^{d+1}w\sqrt{g(w)}K_d(w,\vec{z})
\P^i\phantom{}_k\phantom{}^l\phantom{}_j\tilde{T}^k_l(w).
\ee
Substituting the above expressions for the source into the canonical
momenta and differentiating w.r.t. the scalar source we arrive at the 
 two-point functions 
\bea
\P^i\phantom{}_k\phantom{}^l\phantom{}_j\langle T^k_l(\vec{z})
\mathcal{O}(\vec{w})\rangle=-2(d+1)c
\P^i\phantom{}_d\phantom{}^d
\phantom{}_j I(\vec{z},\vec{w}),\\ \NO\\
\p^l\phantom{}_j\frac{\partial_k}{\square}\langle T^k_l(\vec{z})
\mathcal{O}(\vec{w})\rangle=\frac{\p^k\phantom{}_j}{\square}
\left(\<\co(\vec{z})\>_B\partial_k\d^{(d)}(\vec{z}-\vec{w})\right),\\ \NO\\
\p^l\phantom{}_k\langle T^k_l(\vec{z})
\mathcal{O}(\vec{w})\rangle=-\frac{\partial^k}{\square}
\left(\<\co(\vec{z})\>_B\partial_k\d^{(d)}(\vec{z}-\vec{w})\right),\\ \NO\\ 
\langle T^i_i(\vec{z})\mathcal{O}(\vec{w})\rangle=0,\phantom{more space here}
\eea
where the projection operators are acting on $\vec{z}$ and
\be\label{I_integral}
I(\vec{z},\vec{w})=c_d^2\int d^dx dx_o x_o^{2d+1}
\frac{x_d}{(x_o^2+x_d^2)^{(d+3)/2}}\frac{1}{[x_o^2+(\vec{x}-\vec{z})^2]^d}
\frac{1}{[x_o^2+(\vec{x}-\vec{w})^2]^d}.
\ee
The  last three correlators can be easily seen as a consequence of the Ward identities
(\ref{Ward1}) and (\ref{Ward2}), which is a non-trivial consistency check 
of our calculation. In fact, since the vevs are exact, these two-point 
functions must also be exact in $c$, although our calculation of
the two-point functions was done only to leading order in $c$.

The computation of the remaining of the 2-point functions
$\langle T_{ij}(x) T_{kl}(y)\rangle$ and  $\langle \co(x) \co(y)\rangle$ 
requires an analysis to order $c^2$. This computation is rather complex
since the background metric receives a correction at this order. This 
means that we need to linearize the bulk field equations around the 
corrected solution (\ref{backJ}). The latter, however, is inhomogeneous
and this complicates the analysis. Nevertheless,  the conformal invariance 
of the boundary theory completely determines the two point function of the
scalar operator, while the two-point function of the stress tensor is
determined up to a scalar function (except for $d=2$ where it is 
fully determined) \cite{McAvity&Osborn}. It would be interesting to check that 
the holographic calculation reproduces these two-point functions as well.
 
\subsection{Janus two-point functions vs boundary CFT} 

Let us now take a closer look at the structure of the Janus 
two-point functions. We would like to show that the 2-point functions
are of the form required by conformal invariance for a CFT on a space with 
a wall at $z_d=0$.\footnote{We are
grateful to the authors of \cite{Freedman} for pointing us to the work of McAvity and
Osborn and for prompting us to check that our calculation is
consistent with their results.} The subgroup of the conformal 
group $O(1,d+1)$ that leaves $z_d=0$ invariant is $O(1,d)$.
This is precisely the isometry group of 
the Janus metric. McAvity and Osborn \cite{McAvity&Osborn} have
given explicitly the form of this two-point function in such
a CFT. It is given by
\be
\langle T^i_j(\vec{z})\mathcal{O}(\vec{w})\rangle=-{\rm sgn}(z_d)c\frac{2^{d-1}
d^2\G\left(\frac d2\right)}{(d-1)\p^{d/2}}\left(\frac{v}{\vec{s}^2}
\right)^{d}\left(X^i X_j-\frac1d\d^i_j\right),
\ee
where\footnote{We use $\x$ here to conform with the notation of
\cite{McAvity&Osborn}. This should not be confused with
the argument of the bulk-to-bulk propagator used earlier.}
\be\label{Osborn_notation}
v^2=\frac{\x}{\x+1},\phantom{more} \x=\frac{\vec{s}^2}{4z_dw_d},
\phantom{more} \vec{s}=\vec{z}-\vec{w},
\ee
and
\be
X_i=z_d\frac{v}{\x}\partial_i\x=v\left(\frac{2z_d}{\vec{s}^2}s_i-n_i
\right),
\ee
with $n_i=\d_{id}$.  The normalization of the two-point function
is fixed by the normalization of the vev of the scalar operator
\cite{McAvity&Osborn}. It is clear from (\ref{vevsign}) that this normalization
has opposite signs for $z_d>0$ and $z_d<0$, which is the origin of the 
${\rm sgn}(z_d)$ factor. This expression applies for $z_d w_d>0$, i.e. both
points on the same side of the wall, but not for $z_dw_d<0$. The holographic
expression (\ref{holographic2ptfunction}) applies to both cases, however. 
Under conformal transformations that
leave the hyperplane $z_d=0$ invariant we have
\be
\vec{s}^2\rightarrow \frac{\vec{s}^2}{\Om(\vec{z})\Om(\vec{w})},
\phantom{more} z_d\rightarrow \frac{z_{d}}{\Om(\vec{z})},\phantom{more}
w_d\rightarrow \frac{w_{d}}{\Om(\vec{w})}.
\ee
It follows that $\x$ is a conformal invariant while $X_i$ transforms as 
a vector. In particular, under inversion $\vec{z}\rightarrow \vec{z}/
\vec{z}^2$, $\vec{w}\rightarrow \vec{w}/\vec{w}^2$,
\be
X_i\rightarrow I_{ij}(\vec{z})X_j,
\ee
where $I_{ij}(\vec{z})=\d_{ij}-2\frac{z_i z_j}{\vec{z}^2}$. It is
easy then to see that the two-point function given above transforms
correctly under inversion, namely
\be
\langle T^i_j(\vec{z'})\mathcal{O}(\vec{w'})\rangle=\vec{z}^{2d}
\vec{w}^{2d}I^i_k(\vec{z})I^l_j(\vec{z})
\langle T^k_l(\vec{z})\mathcal{O}(\vec{w})\rangle.
\ee
One can show in general, using the fact that the background has the correct 
isometries, that the holographic 2-point functions transform as they should.
Since the results of \cite{McAvity&Osborn} follow from the same 
symmetries, this argument shows that our results are consistent with
that of \cite{McAvity&Osborn}.
It is, however, a rather 
non-trivial exercise to explicitly demonstrate that the correlator 
is of the form given in \cite{McAvity&Osborn}, mainly because of the 
integral representation of the transverse-traceless part of the correlator.
The integral that appears in the transverse traceless 
part of the holographic two-point function is not easy to evaluate 
in general, and evaluating the projection operator acting on it is not
straightforward either. This makes a direct comparison of the two 
results rather non-trivial. Instead, we will expand both results in
a short distance expansion and compare them term by term. We do this
for the first three orders in the expansion and we find complete 
agreement. 

To facilitate the comparison we first expand the above result of 
McAvity and Osborn. Of course, this expansion is valid only when $z_dw_d>0$, 
which is also the condition for the validity of the McAvity-Osborn expression.
After some algebra we get
\bea
\langle T^i_j(\vec{z})\mathcal{O}(\vec{w})\rangle & = &
-c\frac{2^{d-1}
d^2\G\left(\frac d2\right)}{(d-1)\p^{d/2}}\frac{2w_d}{|2w_d|^{d+1}}
\frac{1}{(\vec{s}^2)^{d/2}}\left\{\frac{s^is_j}{\vec{s}^2}-\frac1d\d^i_j\right. 
\NO \\ 
\left.\right.&\left.\right. &\left.-\frac{1}{2w_d}\left(n^is_j+n_js^i-\vec{n}
\cdot\vec{s}\d^i_j+(d-2)
\vec{n}\cdot\vec{s}\frac{s^is_j}{\vec{s}^2}\right)\right.\NO \\
\left.\right.&\left.\right.&\left.
+\frac{1}{(2w_d)^2}\left[\frac12\left(d(d-2)(\vec{n}\cdot\vec{s})^2-
(d+2)\vec{s}^2\right)\frac{s^is_j}{\vec{s}^2}+2d(\vec{n}\cdot\vec{s})
n^{(i}s_{j)}+\vec{s}^2n^in_j\right.\right.\NO\\
\left.\left.\right.\right.&\left.\left.\right.\right.& \left.\left.
-\frac12\left((d+2)(\vec{n}\cdot\vec{s})^2-\vec{s}^2\right)\d^i_j\right]
+\mathcal{O}(s^3)\right\}.
\eea
The holographic result can be easily evaluated to this order too. First,
using 
\be
\d^{(d)}(\vec{s})=-\frac{\G\left(\frac d2\right)}{2(d-2)\p^{d/2}}\square
\frac{1}{(\vec{s}^2)^{\frac{d-2}{2}}},
\ee
we find that the longitudinal part of the holographic two-point function
reproduces precisely the first two orders of the short distance expansion 
of the McAvity and Osborn result. The transverse traceless part is then
evaluated by acting with the projection operator on
\be\label{I_shortdistanceexp}
I(\vec{z},\vec{w})=\frac{\G\left(\frac d2-1\right)d^2}{8(d+1)\p^{d/2}}
\frac{w_d}{|w_d|^{d+3}}\frac{1}{(\vec{s}^2)^{\frac{d-2}{2}}}
\left(1+\mathcal{O}(s)\right),
\ee 
and it reproduces exactly the third order term. Some details of this 
calculation are presented in appendix \ref{appendix_shortdistexp}. Therefore,
at least to this order in the short distance expansion, we have
shown that the holographic two-point function is exactly what one
expects for a CFT with a wall at $z_d=0$.

\section{Conclusions}
\setcounter{equation}{0}

We discuss in this paper the computation of correlation functions for
holographic RG flows. The computation was done within the Hamiltonian 
framework developed in \cite{PS1}. A central point in our discussion
is that the analysis is focused on the canonical momenta which 
are associated with regularized correlation functions
and not the on-shell action. Furthermore, the renormalization 
procedure is set up such that one only computes the
contribution of counterterms to correlators under discussion
rather than first computing the general counterterms
and then specializing. In particular, to renormalize $n$-point functions
we only need to obtain asymptotic solutions to $(n{-}1)$-order
in fluctuations. For instance, for 2-point functions a linearized analysis is
sufficient. This leads to significant reduction of labor compared to 
previous works.

In the literature the analysis of 
fluctuations and of renormalization were often performed
in different coordinate systems. This was due to the fact that 
renormalization was heavily based in the universal form of
AdS asymptotics which required the use of a  particular coordinate
system, the Fefferman-Graham coordinate system. The fluctuation
equation however is often more easily solvable in other coordinates.
In order to combine the renormalization results 
with the solution of the fluctuation one needs the transformation
between the two coordinate systems.
This is straightforward to obtain, at least asymptotically, which is 
the only thing needed, but it adds to the complexity of the method.
In this work we also avoid this complexity, since the asymptotic
analysis is done directly at the level of the fluctuation
equations. Moreover, it is not necessary to use the Fefferman-Graham
coordinates anymore since the asymptotic expansion is now done 
by organizing the asymptotic solution in terms
of eigenfunctions of the dilatation operator \cite{PS1}.  

To illustrate the method we discussed both flat and AdS-sliced domain 
walls. We reduced the computation of 2-point functions that involve the stress energy
tensor and the operator dual to the active scalar for the most general
flat domain walls driven by a ``superpotential'' $W$ to the solution of two 
second order linear ODEs or (equivalently) to 
two first order non-linear ODEs. With the new method 
it is easy to carry out the renormalization in general (and we discussed how to do this)
but it is even easier -- almost trivial --  to carry out the procedure in each specific 
case. For comparison purposes, we discussed all details for the two cases mostly 
studied in the literature, the GPPZ and CB flows. The new method is comparable in 
efficiency
to the ``old approach'' where one extracts the correlator from the leading non-analytic
part of the linearized solution, but it does not suffer from any of its drawbacks.

As a new example, we discussed the computation of correlation functions of the 
Janus solution. This is a regular non-supersymmetric but stable AdS-sliced
domain wall solution. The main difficulty in this example is that the boundary 
has a corner. However, we showed that there exists a Fefferman-Graham
coordinate system which is well-defined everywhere in the neighborhood of the
boundary except on the corner. This allows 
the vev's of the dual operators to  be read off. We found that the expectation value 
of the stress-energy tensor is zero and the 
expectation value of the operator dual to the active scalar is non-zero.
The expectation value is non-homogeneous and blows up at the location of the corner,
which plays the role of a `wall' in the boundary CFT. 
We computed then the 2-point functions 
that receive a contribution to leading order in $c$, 
i.e. $\langle T_{ij}(x) \co(y)\rangle$. 
Ward identities relate (part of) 
$\langle T_{ij}(x) \co(y)\rangle$ to the vev $\langle \co \rangle$. 
It is a nice check of both the value of the vev and of the 2-point function 
that the Ward identity is indeed satisfied. Both the vev and the 2-point function
we computed are of the form implied by conformal invariance for a CFT
on $\mathbf{R}^d$ with a wall at $z_d=0$.

We discussed here only 2-point functions. Higher point functions have been 
discussed in the context of holographic renormalization 
in \cite{Skenderis:2002wp,3pt}. The new method can be straightforwardly 
applied in such cases, essentially trivializing the issue of 
renormalization. Another direction in which the new method holds promise in 
delivering new
results is the case of solutions that are not asymptotically AdS
such as the Klebanov-Strassler solution \cite{Klebanov:2000hb}.

%\section*{Acknowledgments} 

\acknowledgments 

We would like to thank D. Freedman for informing us about \cite{Freedman}.
KS is supported by NWO.

\begin{appendix}

\section*{Appendix}

\section{Conformal compactification of the Janus solution}
\label{Jage}

In order to determine the conformal compactification of the Janus geometry we 
introduce a new radial coordinate
\be\label{zcoord}
z=\dot{A}=\pm\sqrt{1-u+bu^d}.
\ee
This coordinate has the range $-1\leq z\leq 1$ for {\em any} value
of $b$ and the $u=0$ region is mapped to $z=\pm 1$. $u$ can be determined 
as a function of $z$ by solving the algebraic equation
\be
u-bu^d=1-z^2
\ee
as a power series in $b$. The relevant solution is the smallest real positive root
which is given for arbitrary $d$ and to all orders in $b$ by
\bea\label{ucoordinate}
u(z;b,d)=(1-z^2)\sum_{n=0}^{\infty}\frac{\G(nd+1)}
{\G(n+1)\G(n(d-1)+2)}b^n(1-z^2)^{n(d-1)}=\phantom{more space}\\ \NO
\phantom{}_{(d-1)}F_{(d-2)}\left[\left(\frac{1}{d},\frac{2}{d},\cdots,
\frac{d-1}{d}\right),\left(\frac{2}{d-1},\frac{3}{d-1},\cdots,
\frac{d-2}{d-1},\frac{d}{d-1}\right),\frac{d^d}{(d-1)^{d-1}}b
(1-z^2)^{d-1}\right],
\eea
where $\phantom{}_pF_q$ is the generalized hypergeometric function.
Note that the bound of the $u$ coordinate mentioned above is just $u_o=u(0;b,d)$.
It is not possible to express $u(z)$ in terms of elementary functions
except for the cases $d=2$ and $d=3$. We have respectively,
\be
u(z;b,2)=\frac{1}{2b}\left(1-\sqrt{1-4b(1-z^2)}\right), \quad
u(z;b,3)=\frac{2}{\sqrt{3b}}\sin\left[\frac13\arcsin\left(
\frac32\sqrt{3b}(1-z^2)\right)\right].
\ee

If we now write the (Euclidean) $AdS_d$-slice metric in global coordinates 
and set $z=\sin\theta$, the metric (\ref{exactbackgroundmetric}) becomes
\be
ds_B^2=\frac{1}{u(\sin\theta)\cos^2\l}\left[d\l^2+\cos^2\l\left(1+(2d-1)b
(\cos^2\theta)^{d-1}+\mathcal{O}(b^2)\right)d\theta^2+d\t^2+\sin^2\l d
\Om_{d-2}^2\right],
\ee
where $0\leq\l\leq\p/2$ and $-\p/2\leq\theta\leq\p/2$. A few comments are in 
order here. First, the transformation (\ref{ucoordinate}) implies that 
every power of $b$ in the coefficient of $d\theta^2$ comes with a factor of
$(\cos^2\theta)^{d-1}$ and hence the metric inside the square brackets is 
non-singular for any $\theta$. Second, note that the $(\l,\theta)$ part of 
the metric can be transformed into the standard metric on $\mathbf{S}^2$ by
introducing the angular coordinate
\be
\m=\int_0^{\sin\theta}\frac{dz}{\sqrt{u(z)}\left(1-bdu(z)^{d-1}\right)}=
\theta+\left(d-\frac12\right)b\sin\theta F\left(\frac12,\frac32-d;\frac32;
\sin^2\theta\right)+\mathcal{O}(b^2).
\ee
This is precisely the angular coordinate introduced in \cite{BGH,FNSS} and
it takes values in $[-\m_o,\m_o]$, where $\m_o\geq\p/2$ is given in equation
(B.8) of \cite{FNSS}. Because of the excess angle the compact metric has a 
corner at $\l=\p/2$, as is discussed in \cite{FNSS}.

\section{Fefferman-Graham coordinates for Janus metric}
\label{appendix_FG}
To construct the Fefferman-Graham metric we start with 
(\ref{exactbackgroundmetric}) and the coordinate transformation (\ref{ucoordinate}) 
and write the $AdS_d$-slice metric in the upper-half plane coordinates. Then  
\bea
ds^2_B=\frac{dz^2}{(1-z^2)^2}\left[1+2(d-1)b(1-z^2)^{d-1}+
\mathcal{O}(b^2)\right]+\NO \\
\frac{1}{1-z^2}\left[1-b(1-z^2)^{d-1}+\mathcal{O}(b^2)\right]
\frac{1}{\tilde{z}_o^2}(d\tilde{z}_o^2+dz_a^2),
\eea
where  $a=1,\ldots,d-1$. For $b=0$ the coordinate transformation
\be\label{coordtrans}
z=\frac{z_d}{\sqrt{z_o^2+z_d^2}},\,\,\,\,\,\tilde{z}_o=\sqrt{z_o^2+z_d^2}
\ee
brings this metric into the upper half plane metric with radial
coordinate $z_o$. To determine the Fefferman-Graham form of the Janus
metric we need to obtain appropriate $b$-dependent corrections to this 
transformation. We can
determine these as a Taylor series
in $b$ by introducing two arbitrary functions at each order in
$b$ and solving the differential equations that result by requiring
that the transformed metric is of the Fefferman-Graham form. The unique
transformation which ensures that the metric remains asymptotically
AdS independent of $b$ is to linear order in $b$
\be\label{coordexpansion}
z=\frac{z_d}{\sqrt{z_o^2+z_d^2}}+bf_1(x)+\mathcal{O}(b^2),\,\,\,\,\,
\tilde{z}_o=\sqrt{z_o^2+z_d^2}+bz_of_2(x)+\mathcal{O}(b^2),
\ee
where $x\equiv z_d/z_o$ and
\bea
f_1(x)=\frac{x}{2(1+x^2)^{3/2}}\left[\left(1-\frac{1}{2d}\right)
\frac{1}{x^{2d}}F\left(d,d;d+1;-\frac{1}{x^2}\right)+\phantom{more space}
\right.\NO \\ 
\left.\phantom{more space}\frac{1}{2(d+1)x^{2(d+1)}}F\left(d+1,d;d+2;-
\frac{1}{x^2}\right)+\frac{1}{(1+x^2)^{d-1}}\right], \\ \NO \\ 
f_2(x)=\frac{1}{2\sqrt{1+x^2}}\left[\frac{1+2dx^2}{2dx^{2d}}
F\left(d,d;d+1;-\frac{1}{x^2}\right)+\phantom{more space}
\right.\NO \\ 
\left.\phantom{more space}\frac{1}{2(d+1)x^{2d}}F\left(d+1,d;d+2;-
\frac{1}{x^2}\right)-\frac{1}{(1+x^2)^{d-1}}\right].
\eea

The metric then takes the from
\be\label{backJ}
ds_B^2=\frac{1}{z_o^2}\left[dz_o^2+(1+bc_3(x)+\mathcal{O}(b^2))dz_d^2+
(1+bc_4(x)+\mathcal{O}(b^2))dz_a^2\right], 
\ee
where
\bea
&&c_3(x)=\frac{(2d-1)}{2dx^{2d}}F(d,d;d+1;-\frac{1}{x^2})+
\frac{1}{2(d+1)x^{2(d+1)}}F(d+1,d;d+2;-\frac{1}{x^2})-
\frac{1}{x^2(1+x^2)^{d-1}},\NO \\ 
&&c_4(x)=-\frac{1}{2dx^{2d}}F(d,d;d+1;-\frac{1}{x^2}).
\eea
Note that the derivatives of
these functions have a much simpler form:
\be
c_3'(x)=\frac{1}{x^3(1+x^2)^d}\left[-1-(2d-1)x^2+2\frac{1+(d+2)x^2}
{(1+x^2)^2}\right],\quad
c_4'(x)=\frac{1}{x(1+x^2)^d}.
\ee
The metric in (\ref{backJ}) is manifestly invariant under 
translations and rotations of the $z_a$ coordinates and scale
transformations (the $x$ coordinate is invariant under scale
transformations). The original metric (\ref{exactbackgroundmetric})
however was invariant under the larger group $O(1,d)$ 
associated with the AdS slice metric.
We now show  that the metric (\ref{backJ})
is also invariant under a discrete inversion 
isometry to order $b$ which enhances the isometry group 
to the full $O(1,d)$. Actually, we will see that
the inversion symmetry can be used to obtain the 
Fefferman-Graham form of the metric to all orders in $b$. 

Let us write the AdS slice metric in (\ref{exactbackgroundmetric}) in
the upper half plane coordinates so that
\beq
ds_B^2=\frac{du^2}{4u^2(1-u+bu^d)}+\frac{1}{u \tilde{z}_o^2}(d\tilde{z}_o^2+dz^a dz^a).
\eeq 
This form is invariant under the discrete isometry
\be
\tilde{z}_o\rightarrow \frac{\tilde{z}_o}{\tilde{z}_o^2+z_a^2},\phantom{more space}
z_a\rightarrow \frac{z_a}{\tilde{z}_o^2+z_a^2}.
\ee
We now bring this metric into the Fefferman-Graham form by means of
a coordinate transformation\footnote{Note that $u=u(z;b)$ is given in
eq. (\ref{ucoordinate}).}
\be\label{coordtrans_nonpert}
z=s(x;b),\phantom{more space} \tilde{z}_o=z_o t(x;b),
\ee
where $x=z_d/z_o$. We point out that this is precisely the form of 
the coordinate transformation (\ref{coordexpansion}), but we now 
treat the $b$-dependence non-perturbatively. This allows us to
express the above discrete isometry in terms of the Fefferman-Graham 
coordinates $z^\m=(z_o,z_a,z_d)$. We find
\be
z^\m\rightarrow \frac{z^\m}{z_o^2t(x;b)^2+z_a^2}.
\ee 
Now, the Fefferman-Graham metric 
(\ref{backJ}) takes the form
\beq
ds_B^2=\frac{1}{z_o^2}\left[dz_o^2+\l(x;b)dz_d^2+
\m(x;b)dz_a^2\right].
\eeq
The requirement that this is invariant under inversion uniquely
determines the functions $\l(x;b)$ and $\m(x;b)$ in terms of
$t(x;b)$. Namely we find the exact FG metric
\beq\label{exactFGmetric}
\framebox[\width]{
\begin{minipage}{5.in}
\begin{eqnarray*}
ds_B^2=\frac{1}{z_o^2}\left[dz_o^2+\frac{\pa_x t}{x(t-x \pa_x t)}dz_d^2+
\frac{1}{t(t-x \pa_x t)}dz_a^2\right].\\
\end{eqnarray*}
\end{minipage}}
\eeq
Requiring further that this is equal to the Janus metric above
uniquely fixes the transformation functions $s(x;b)$ and $t(x;b)$.
In particular we obtain the system of coupled equations
\be\label{s_t_equations}
u=1-x\frac{\pa_x t}{t},\phantom{more} (\pa_x s)^2=
\frac{1}{x^2}u^2(1-u)(1-bdu^{d-1})^2,
\ee
where $u=u(s(x))$ is given by (\ref{ucoordinate}). 

In order to solve these equations we use (\ref{zcoord}) to trade
$s(x)$ for $u(x)$ in the second equation, which gives
\be\label{u_x_relation}
\framebox[\width]{
\begin{minipage}{5.in}
\begin{eqnarray*}
\int^{u(x)}\frac{du'}{u'\sqrt{(1-u')(1-u'+bu'^d)}}=
-\log x^2.\\
\end{eqnarray*}
\end{minipage}}
\ee
The sign and the integration constant are chosen so that 
$u(x)\sim 1/x^2$ as $x\to\infty$, independent of $b$.
Unfortunately it seems rather difficult to do this integral
explicitly for arbitrary dimension $d$. Instead, one can expand the integrand 
in $b$ and integrate term by term. This gives
\be\label{u_x_expansion}
u(x)=\frac{1}{1+x^2}+\frac{b}{2d}\frac{x^2}{(1+x^2)^{d+2}}
F(d,2;d+1;\frac{1}{1+x^2})+\mathcal{O}(b^2).
\ee
The transformation functions $s(x;b)$ and $t(x;b)$ are now
determined from 
\be
\framebox[\width]{
\begin{minipage}{5.in}
\begin{eqnarray*}
s(x) & = & 1-u(x)+bu(x)^d \\ \NO \\
t(x) & = & \exp\left[\frac12\int_{u(x)}^1\frac{du'}{u'}
\left(\frac{1-u'}{1-u'+bu'^d}\right)^{1/2}\right].\label{t}\\
\end{eqnarray*}
\end{minipage}}
\ee
Inserting the above expansion for $u(x)$ we reproduce 
(after some manipulation of the hypergeometric functions)
precisely the coordinate transformation (\ref{coordexpansion}).  

Moreover, inserting (\ref{u_x_expansion}) in  
\be
\framebox[\width]{
\begin{minipage}{5.in}
\begin{eqnarray*}
\phi_B(x)=\phi_o+c\int_0^x\frac{dx'}{|x'|}u(x')^{d/2}\sqrt{1-u(x')}\\
\end{eqnarray*}
\end{minipage}}
\ee 
gives
\be\label{scalarexpansion}
\f_B(x)=\f_o+c c_5(x)+\mathcal{O}(c^3),
\ee
where $\f_o$ is a constant and
\be
c_5(x)=\frac{x}{\sqrt{1+x^2}}F\left(\frac12,1-\frac d2;\frac32;
\frac{x^2}{1+x^2}\right).
\ee
Again, this has a simple derivative:
\be
c_5'(x)=\frac{1}{(1+x^2)^{(d+1)/2}}.
\ee
Notice that as $z_o \to 0$ with all other coordinates fixed 
(i.e. as we approach the 
conformal boundary) $c_5(x) = {\rm sgn}(z_d)$ while higher order terms
do not contribute.  So at the boundary
\be \label{coupl}
\f_B(z_d)=\f_o + {\rm sgn}(z_d)c.
\ee
This implies that the coupling of the dual operator is different on the 
two sides of the wall.

Finally, let us examine the range of validity of 
the coordinate transformation (\ref{coordtrans_nonpert}). 
The Jacobian of the transformation is equal to $J=t \pa_x s$. 
Now $J=0$ implies $\pa_x s=0$ since 
$t(x)$ is positive definite, as can be seen from (\ref{t}).
It follows that   
the coordinate transformation breaks down at $u=1$. Note
that the zero of $(1-bdu^{d-1})$ occurs at $u=1/(bd)^{1/(d-1)}>1$, where
the inequality follows from (\ref{brange}). We
conclude that the Fefferman-Graham coordinates are valid in the 
range $0<u<1$ although, in general $0\leq u\leq u_o$ with $u_o\geq 1$.
Recall that in general the Fefferman-Graham coordinate system is only 
guaranteed to exist in a neighborhood of the boundary, and here we 
see an explicit illustration of this. 

Recall that the Fefferman-Graham coordinate system \cite{FG} is obtained
as follows (see section 3 of \cite{Skenderis:2002wp} for a review). 
One considers Gaussian normal coordinates centered at the  boundary 
and the radial coordinate is 
identified with the affine parameter of the geodesics emanating perpendicularly 
from the boundary. Clearly the
region of validity of this coordinate system depends on the behavior
of the radial geodesics. We therefore need to analyze such geodesics,
and we will do this in the $(u,\tilde{z}_o,z_a)$ coordinate system
which is well-defined everywhere.

\begin{figure}
\begin{center}
\scalebox{0.7}{\rotatebox{-0}{\includegraphics{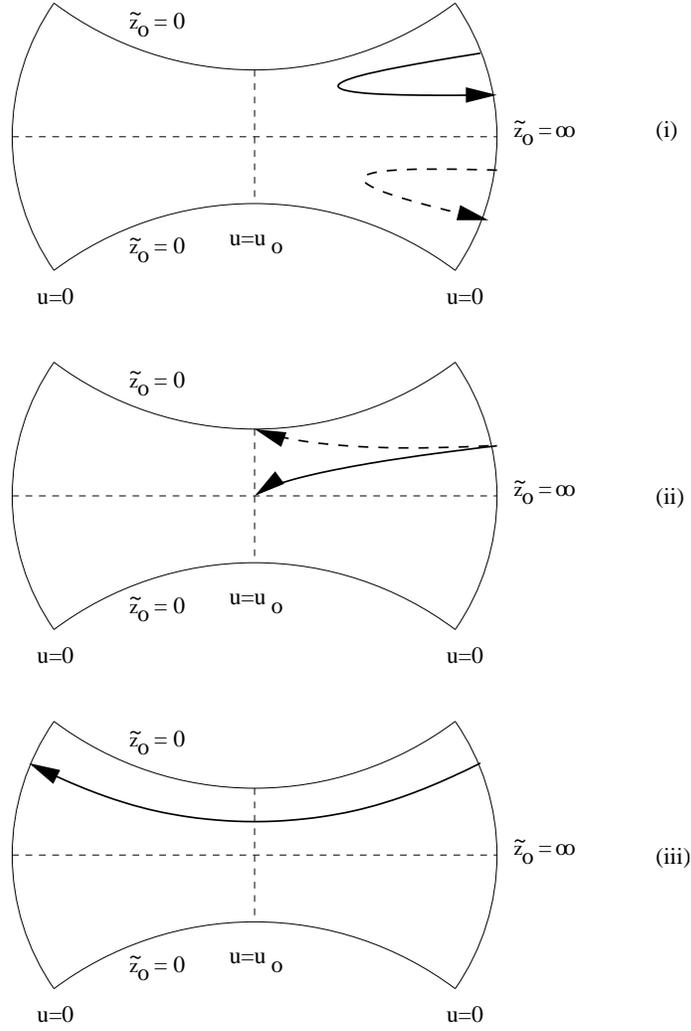}}}
\end{center}
\caption[]{\small The qualitatively different radial geodesics of the Janus geometry:
(i) $a_2<u_o$ with $a_1>0$ (solid arrow) or $a_1<0$ (broken arrow), (ii)
$a_2=u_o$ with $a_1>0$ (solid arrow) or $a_1<0$ (broken arrow) and (iii) $a_1=0$. 
These qualitative features are insensitive to the value of $b$. }
\label{geodesics}
\end{figure}

One easily shows that there are geodesics with $z_a$ constant.
The geodesic equations for the remaining coordinates
lead to the following two equations
\bea
\frac{d\log\tilde{z}_o}{d\t}=a_1u,\\ \NO \\
\ddot{u}-\left(\frac1u+\frac{-1+bdu^{d-1}}{2(1-u+bu^d)}\right)\dot{u}^2+
2a_1^2u^2(1-u+bu^d)=0,
\eea
where $a_1$ is an integration constant. If $a_1\neq 0$ the second equation can be 
integrated once to get 
\be 
\dot{u}=\pm 2a_1u\sqrt{(a_2-u)(1-u+bu^d)}
\ee 
for some constant $0<a_2\leq u_o$. If $a_1=0$ one gets instead
\be 
\dot{u}= \pm a_3u\sqrt{1-u+bu^d},
\ee 
where $a_3$ is again a constant. Now, depending on the values of the
parameters $a_1$ and $a_2$, we can identify three qualitatively different
types of geodesics as shown in fig.\ref{geodesics}. 

Consider now the radial geodesics defined by $\dot{z}_d=\dot{z}_a=0$
in the Fefferman-Graham coordinates, where the dot stands for the derivative
w.r.t. the affine parameter $\t=\log z_o$. Since $z_d={\rm constant}$ along these
geodesics we will take $\t=\log(z_o/|z_d|)=-\log|x|$ for later convenience.
The transformation (\ref{coordtrans_nonpert}) immediately gives
\be
\frac{d\log\tilde{z}_o}{d\t}=u,
\ee
while (\ref{u_x_relation}) implies
\be
\dot{u}=2u\sqrt{(1-u)(1-u+bu^d)}.
\ee
The Fefferman-Graham radial geodesics therefore correspond to radial  geodesics 
 with $a_1=a_2=1$. In particular, they are geodesics of type (i) if $b>0$ but they 
are type (ii) if $b=0$. This is an important qualitative difference between the 
FG coordinates for the Janus geometry and pure AdS. This is in fact why
the FG coordinates cover the whole of AdS but only part of the Janus geometry. 

\begin{figure}
\begin{center}
\scalebox{0.7}{\rotatebox{-0}{\includegraphics{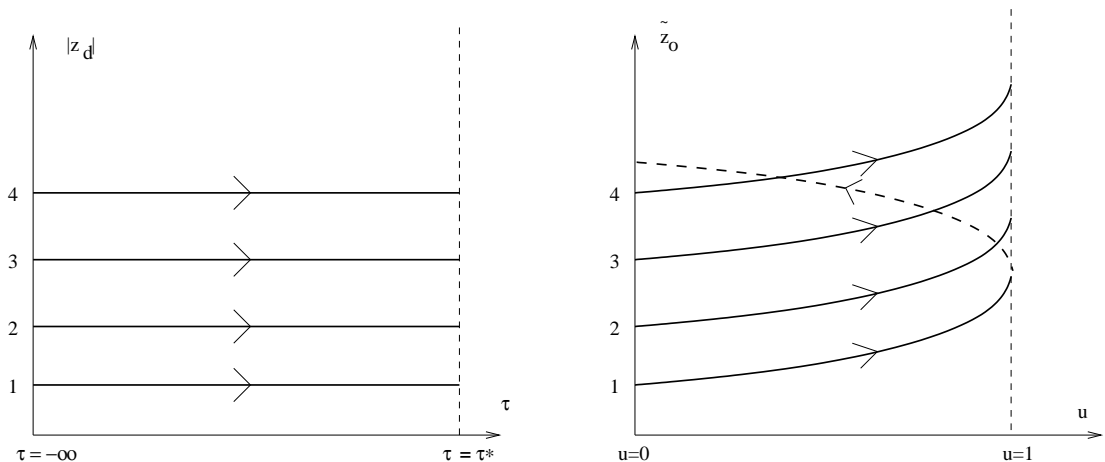}}}
\end{center}
\caption[]{\small  A radial geodesic in the Fefferman-Graham coordinates (left)
is not defined beyond $u=1$ due to the failure of the coordinate system at 
this point. In the coordinate system which extends beyond $u=1$ this geodesic
bounces back at $u=1$. Only the branch before the bounce corresponds to
the geodesic on the left. }
\label{figure}
\end{figure}

It is now clear why the FG coordinate system for $b>0$ breaks down at $u=1$.
Namely, the radial FG coordinate corresponds to geodesics
which do not reach beyond $u=1$. If one continues to affine parameter
values greater than $\t*$, where $u(\t*)=1$, the geodesics bounce back and 
they cannot be used to define a coordinate system since they doubly cover the
region $u<1$ as is shown in  fig.\ref{figure}.
Therefore the FG coordinates are well-defined for affine parameter
values $\t<\t*$. This means that $|x|=e^{-\t}$ must be  bounded below.  Another
way to see this is to observe that (\ref{u_x_relation}) implies 
that $x^2$ is a monotonically
decreasing function of $u$. Hence the upper bound $u<1$ on $u$ implies a 
lower bound on $x^2$. Setting $u=1$ in (\ref{u_x_expansion}) and solving for
$x$ to leading order in $b$ we find\footnote{One must be careful since the
hypergeometric function $F(d,2;d+1;\frac{1}{1+x^2})$ is singular at $x^2=0$.
See eq. 15.3.12 in \cite{A&S}.} $|x|>x_o=b/\sqrt{2}+\mathcal{O}(b^2)$. 
Therefore $x=0$ is {\em not} part of the manifold and hence 
 the metric (\ref{backJ}) (and (\ref{exactFGmetric})) is {\it non-singular} 
in the region it is well-defined.  (To cover the entire spacetime
one would have to use another coordinate patch that covers the deep
interior region $1\leq u \leq u_o$, but this is irrelevant for our
holographic computations.) Notice that the bound on $x$ translates into 
$|z_d|>x_oz_o$ and so
only at the boundary $z_o\to 0$ does $z_d$ cover the entire real line.
More crucially, the bound $z_o<|z_d|/x_o$ means that the radial coordinate,
$z_o$, is well-defined everywhere {\em except} at $z_d=0$, which precisely 
corresponds to the corner where the two-halves of the boundary meet. This
can also be seen directly from the properties of type (i) geodesics. 
Since $\tilde{z}_o\to |z_d|$ as $\t\to -\infty$ with $z_d$ 
constant along the geodesics, we have
\be
\tilde{z}_o(\t)=|z_d|\exp\int_{-\infty}^\t d\t'u(\t').
\ee 
So for $b>0$ the FG geodesics hit again the boundary $u=0$ at $\tilde{z}_o=|z_d|\a$,
where $\a=\exp\int_0^1\frac{du}{\sqrt{(1-u)(1-u+bu^d)}}$. At  $z_d=0$ 
these geodesics become degenerate and they stay along $\tilde{z}_o=0$. Since
the entire $\tilde{z}_o=0$ subspace is mapped to $(z_o,z_d)=(0,0)$ in the
FG coordinates, the FG geodesics do not leave the origin once $z_d=0$. Hence,
the FG radial coordinate is not defined at $z_d=0$. 

Finally let us discuss the possibility to use the geodesics of type (ii) in order to 
define FG coordinates. This is also a natural choice as these geodesics
are the obvious generalization of the pure AdS case.
Following radial type (ii) geodesics for the Janus geometry, the exact FG 
metric is
\be
ds^2=\frac{dz_o^2}{z_o^2}+\frac{u_o(u_o-u)}{uz_d^2}dz_d^2+\frac{dz_a^2}{u\tilde{z}_o^2}
\ee
where
\be
\log\left(\frac{z_o^{2/\sqrt{u_o}}}{z_d^2}\right)=\int^u
\frac{du}{u\sqrt{(u_o-u)(1-u+bu^d)}}
\ee
and
\be
\tilde{z}_o^2=z_d^{2u_o}\exp\int_0^u\frac{du}{\sqrt{(u_o-u)(1-u+bu^d)}}
\ee
Asymptotically, i.e. as $z_o\to 0$,
\be
ds^2\sim \frac{1}{z_o^2}\left[dz_o^2+(\tilde{z}_d^2)^{\frac{1}{\sqrt{u_0}}-1}
(d\tilde{z}_d^2+dz_a^2)\right]
\ee
where $\tilde{z}_d = z_d^{u_0}$. So when $u_o=1$, i.e. the pure AdS case,
we get the standard result but for the Janus solution these geodesics lead to a non-flat 
representative of the conformally flat conformal structure. One can perform
the additional change of variables given in (8)-(9) of \cite{Skenderis:2000in}
to change the representative to the flat metric. Notice however that the 
conformal factor is singular at $z_d=0$ so the corresponding coordinate 
transformation is singular there. We thus arrive again at the conclusion that 
the FG coordinates are not well defined at the corner.

We have therefore determined the exact form of the Fefferman-Graham metric for the
Janus geometry and we have shown that is well-defined everywhere except on the 
defect where the two half-boundaries are joined and it is non-singular where 
it is defined. 
Moreover, we have shown that the FG metric possesses an inversion isometry which 
 enhances the isometry group to the full $O(1,d)$ isometry group 
of the original Janus metric. 
This is reflected  
in the fact that the holographic calculation gives a zero vev for the 
stress tensor, which is consistent with a boundary QFT invariant
under conformal transformations leaving the plane $z_d=0$ invariant.

\section{Two-point functions for Janus from scalar momentum}
\label{appendix_2ptfns}
Here we give some details of the calculation of the two-point
functions for the Janus background based on the scalar equation
of motion (\ref{scalarequation}). The non-trivial part of the
calculation consists in casting the source in a form which 
significantly reduces the amount of work required. To this end
we again use the fact that $\phi_B(x)$ is a function of $x=z_d/z_o$
only and it satisfies
\be
\ddot{\phi}_B+d\dot{\phi}_B+e^{-2r}\square \phi_B=\mathcal{O}(b)
\ee
to write
\be
\partial_j\phi_B(x)=\d_{jd}\frac{1}{z_d}\dot{\phi}_B,\phantom{more}
\partial_i\partial_j\phi_B=-(d+1)\d_{id}\d_{jd}
\frac{\dot{\phi}_B}{z_o^2+z_d^2}+\mathcal{O}(b).
\ee
Decomposing $S^i_j$ into irreducible components then (\ref{scalarequation})
becomes
\bea
-\square_g\phi & = & -cz_o\partial_i\left[\frac{1}{(1+x^2)^{\frac{d+1}{2}}}
e^i_d\right]\NO \\ & & +\frac{cx}{(1+x^2)^{\frac{d+1}{2}}}\left\{-
\frac{z_o^2}{2(d-1)}
\square f+\frac{(d+1)}{1+x^2}\left[2\partial_{z_d}\e_d+\frac{d}{d-1}
\left(\frac1d-\frac{\partial_{z_d}^2}{\square}\right)f+
\frac{\partial_{z_d}^2}{\square}S\right]\NO \right.\\ & & \left.
-\frac{z_o^2}{z_d}
\left[\square \e_d-\partial_{z_d}f+\frac12\partial_{z_d}S\right]\right\}
+\mathcal{O}(b).
\eea
Quite remarkably, this can be cast in the form
\be 
-\square_g\phi=-cz_o\partial_i\left[\frac{1}{(1+x^2)^{\frac{d+1}{2}}}
e^i_d\right]-\square_g\tilde{J}_\phi+\mathcal{O}(b),
\ee
where
\be
\tilde{J}_\phi=\frac{cz_o^d}{(z_o^2+z_d^2)^{\frac{d+1}{2}}}\left[\a +\e_d+
\frac12\frac{\partial_{z_d}}{\square}S+\frac{1}{2(d-1)}\left(z_d-d
\frac{\partial_{z_d}}{\square}\right)f\right],
\ee
and $\a$ is a constant. Hence, the inhomogeneous solution is
\be
\phi=\tilde{J}_\phi-c\int d^{d+1}w\sqrt{g(w)}G(\x)\partial_i
\left[\frac{z_o}{(1+x^2)^{\frac{d+1}{2}}}e^i_d\right],
\ee
where $e^i_d$ is given by the zero-order solution
\be
e^i_j(z)=\int d^dy K_d(z,\vec{y})e\sub{0}^i_j(\vec{y}).
\ee
This expression for $\phi$ immediately gives the canonical momentum from 
which we obtain the two-point function by differentiating w.r.t. 
$S\sub{0}^i_j$.\footnote{Note that
\begin{eqnarray*}\langle T_{ij}(x)\mathcal{O}(x')\rangle=\left(-\frac{1}{
\sqrt{g\sub{0}(x')}}\frac{\d}{\d\phi\sub{0}(x')}\right)
\left(-\frac{2}{
\sqrt{g\sub{0}(x)}}\frac{\d}{\d g\sub{0}^{ij}(x)}\right)W
=\left(-\frac{1}{
\sqrt{g\sub{0}(x')}}\frac{\d}{\d\phi\sub{0}(x')}\right)\langle T_{ij}(x)
\rangle \\ =\left(-\frac{2}{
\sqrt{g\sub{0}(x)}}\frac{\d}{\d g\sub{0}^{ij}(x)}\right)\langle 
\mathcal{O}(x')\rangle+\d^{(d)}(x,x')g\sub{0}_{ij}(x)\langle 
\mathcal{O}(x')\rangle,\end{eqnarray*}
so if one starts the computation from the scalar 1-point functions, one
should remember to include the contact term given above to obtain the full
expression.} The result is
\bea\label{holographic2ptfunction}
\langle T^i_j(\vec{z})\mathcal{O}(\vec{w})\rangle & = &
-2(d+1)c
\P^i\phantom{}_d\phantom{}^d
\phantom{}_j I(\vec{z},\vec{w}) \NO \\ &&
-\frac{cd}{|w_d|^{d+1}}\left[ 2\p^{(i}_d\frac{\partial_{j)}}{\square}+
\d^i_j\frac{\partial_d}{\square}+\frac{1}{d-1}\left(
w_d-d\frac{\partial_d}{\square}\right)\p^i_j-\frac1d w_d\d^i_j\right]_w
\d^{(d)}(\vec{z}-\vec{w}).\NO \\
\eea
It is a straightforward exercise to verify that this is equivalent
to the two-point functions given above, as calculated from the 
graviton momentum.

\section{Short distance expansion of the holographic two-point function 
$\langle T^i_j(\vec{z})\mathcal{O}(\vec{w})\rangle$}
\label{appendix_shortdistexp}
For the convenience of the reader we will give here the essential steps 
required to evaluate the short distance expansion of the transverse 
traceless part of the holographic two-point function 
$\langle T^i_j(\vec{z})\mathcal{O}(\vec{w})\rangle$, namely 
$\P^i\phantom{}_d\phantom{}^d
\phantom{}_j I(\vec{z},\vec{w})$, where $I(\vec{z},\vec{w})$ is given
in (\ref{I_integral}).\footnote{Incidentally, this integral transforms 
under inversion as $I(\vec{z'},\vec{w'})=\vec{z}^{2d}\vec{w}^{2d}
I(\vec{z},\vec{w})$ and hence it must be of the form $f(v)/(\vec{s}^2)^d$
for some function $f(v)$, where $v$ is defined in (\ref{Osborn_notation}).
However, we have not succeeded in determining this function so far.}

First, after a shift and rescaling of the integration variables, 
 $I(\vec{z},\vec{w})$ can be written as
\be
I(\vec{z},\vec{w})=\frac{c_d^2}{(\vec{s}^2)^{d/2-1}}
\int_0^\infty dx_o x_o^{2d+1}\int d^dx \frac{w_d+|\vec{s}|x_d}
{\left[x_o^2\vec{s}^2+(w_d+|\vec{s}|x_d)^2\right]^{\frac{d+3}{2}}}
\frac{1}{\left[x_o^2+\vec{x}^2\right]^d} \frac{1}{\left[x_o^2+
(\vec{x}-\hat{s})^2\right]^d},
\ee
where $\hat{s}=\vec{s}/\vec{s}^2$. This form is suitable for a short
distance expansion in $|\vec{s}|$. Each term in the expansion can be 
explicitly evaluated using the standard Feynman parameters technique.
The result to leading order is given in (\ref{I_shortdistanceexp}). 

To evaluate the projection operator on this expression we use
the fact that
\be
\frac{1}{(\vec{s}^2)^\a}=-\frac{1}{2(\a-1)(d-2\a)}
\square\frac{1}{(\vec{s}^2)^{\a-1}},
\ee
for any power $\a\neq d/2$ in order to cancel the $1/\square$ factors 
in the projection operator. It is then straightforward to evaluate
the derivatives in the numerator of the projection operator to obtain
the short distance expansion of the transverse traceless part. The
result is given in section 4.4.
 
\end{appendix}


\begin{thebibliography}{99}

%\cite{Gubser:1998bc}
\bibitem{Gubser:1998bc}
S.~S.~Gubser, I.~R.~Klebanov and A.~M.~Polyakov,
``Gauge theory correlators from non-critical string theory,''
Phys.\ Lett.\ B {\bf 428}, 105 (1998)
[hep-th/9802109].
%%CITATION = HEP-TH 9802109;%%

%\cite{Witten:1998qj}
\bibitem{Witten:1998qj}
E.~Witten,
``Anti-de Sitter space and holography,''
Adv.\ Theor.\ Math.\ Phys.\  {\bf 2}, 253 (1998)
[hep-th/9802150].
%%CITATION = HEP-TH 9802150;%%

\bibitem{dHSS}
S.~de Haro, S.~N.~Solodukhin and K.~Skenderis,
``Holographic reconstruction of spacetime and renormalization in the
 AdS/CFT correspondence,''
Commun.\ Math.\ Phys.\  {\bf 217} (2001) 595
[hep-th/0002230].
%%CITATION = HEP-TH 0002230;%%

\bibitem{howtogo}
M.~Bianchi, D.~Z.~Freedman and K.~Skenderis,
``How to go with an RG flow,''
JHEP {\bf 0108} (2001) 041
[hep-th/0105276].
%%CITATION = HEP-TH 0105276;%%

\bibitem{holren}
M.~Bianchi, D.~Z.~Freedman and K.~Skenderis,
``Holographic renormalization,''
Nucl.\ Phys.\ B {\bf 631} (2002) 159
[hep-th/0112119].
%%CITATION = HEP-TH 0112119;%%

\bibitem{HS}
M.~Henningson and K.~Skenderis,
``The holographic Weyl anomaly,''
JHEP {\bf 9807} (1998) 023
[hep-th/9806087];
%%CITATION = HEP-TH 9806087;%%
M.~Henningson and K.~Skenderis,
``Holography and the Weyl anomaly,''
Fortsch.\ Phys.\  {\bf 48} (2000) 125
[hep-th/9812032].
%%CITATION = HEP-TH 9812032;%%

%\cite{Skenderis:2002wp}
\bibitem{Skenderis:2002wp}
K.~Skenderis,
``Lecture notes on holographic renormalization,''
Class.\ Quant.\ Grav.\  {\bf 19} (2002) 5849
[hep-th/0209067].
%%CITATION = HEP-TH 0209067;%%

\bibitem{fmmr}
D.~Z.~Freedman, S.~D.~Mathur, A.~Matusis and L.~Rastelli,
``Correlation functions in the CFT($d$)/AdS($d+1$) correspondence,''
Nucl.\ Phys.\ B {\bf 546}, 96 (1999)
[hep-th/9804058].
%%CITATION = HEP-TH 9804058;%%

\bibitem{PS1}
I.~Papadimitriou and K.~Skenderis,
``AdS/CFT correspondence and geometry,''
[hep-th/0404176].
%%CITATION = HEP-TH 0404176;%%


%\cite{Kraus:1999di}
\bibitem{Kraus:1999di}
P.~Kraus, F.~Larsen and R.~Siebelink,
``The gravitational action in asymptotically AdS and flat spacetimes,''
Nucl.\ Phys.\ B {\bf 563} (1999) 259
[hep-th/9906127].

%\cite{deBoer:1999xf}
\bibitem{deBoer:1999xf}
J.~de Boer, E.~Verlinde and H.~Verlinde,
``On the holographic renormalization group,''
JHEP {\bf 0008} (2000) 003
[hep-th/9912012];
%%CITATION = HEP-TH 9912012;%%
J.~de Boer,
``The holographic renormalization group,''
Fortsch.\ Phys.\  {\bf 49}, 339 (2001)
[hep-th/0101026].
%%CITATION = HEP-TH 0101026;%%


%\cite{Martelli:2002sp}
\bibitem{Martelli:2002sp}
D.~Martelli and W.~Muck,
``Holographic renormalization and Ward identities with the  
Hamilton-Jacobi method,''
Nucl.\ Phys.\ B {\bf 654} (2003) 248
[arXiv:hep-th/0205061];
%%CITATION = HEP-TH 0205061;%%

%\cite{Fukuma:2002sb}
\bibitem{Fukuma:2002sb}
M.~Fukuma, S.~Matsuura and T.~Sakai,
``A note on the Weyl anomaly in the holographic renormalization group,''
Prog.\ Theor.\ Phys.\  {\bf 104}, 1089 (2000)
[hep-th/0007062];
%%CITATION = HEP-TH 0007062;%%
J.~Kalkkinen and D.~Martelli,
``Holographic renormalization group with fermions and form fields,''
Nucl.\ Phys.\ B {\bf 596}, 415 (2001)
[hep-th/0007234];
%%CITATION = HEP-TH 0007234;%%
J.~Kalkkinen, D.~Martelli and W.~Muck,
``Holographic renormalisation and anomalies,''
JHEP {\bf 0104}, 036 (2001)
[arXiv:hep-th/0103111];
%%CITATION = HEP-TH 0103111;%%
M.~Fukuma, S.~Matsuura and T.~Sakai,
``Holographic renormalization group,''
Prog.\ Theor.\ Phys.\  {\bf 109}, 489 (2003)
[hep-th/0212314];
%%CITATION = HEP-TH 0212314;%%
M.~Banados, A.~Schwimmer and S.~Theisen,
``Chern-Simons gravity and holographic anomalies,''
JHEP {\bf 0405}, 039 (2004)
[arXiv:hep-th/0404245].
%%CITATION = HEP-TH 0404245;%%

\bibitem{BGH}
D.~Bak, M.~Gutperle and S.~Hirano,
``A dilatonic deformation of AdS(5) and its field theory dual,''
JHEP {\bf 0305}, 072 (2003)
[hep-th/0304129].
%%CITATION = HEP-TH 0304129;%%

\bibitem{FNSS}
D.~Z.~Freedman, C.~Nunez, M.~Schnabl and K.~Skenderis,
``Fake supergravity and domain wall stability,''
Phys.\ Rev.\ D {\bf 69}, 104027 (2004)
[hep-th/0312055].
%%CITATION = HEP-TH 0312055;%%

%\cite{Petkou:1999fv}
\bibitem{Petkou:1999fv}
A.~Petkou and K.~Skenderis,
``A non-renormalization theorem for conformal anomalies,''
Nucl.\ Phys.\ B {\bf 561}, 100 (1999)
[arXiv:hep-th/9906030].
%%CITATION = HEP-TH 9906030;%%


\bibitem{Townsend} P.K. Townsend, 
``Positive energy and the scalar potential in higher dimensional 
(super)gravity theories,
Phys. Lett. {\bf 148B} (1984) 55.

\bibitem{ST} K.~Skenderis and P.~K.~Townsend,
``Gravitational stability and renormalization-group flow'',
Phys.\ Lett.\ B {\bf 468}, 46 (1999)
[hep-th/9909070].
%%CITATION = HEP-TH 9909070;%%

\bibitem{GPPZ} L. Girardello, M. Petrini, M. Porrati and A. Zaffaroni,
``The Supergravity Dual of $N=1$ Super Yang-Mills Theory''
Nucl. Phys. {\bf B569} (2000) 451-469, 
[hep-th/9909047].

\bibitem{CB} D.Z. Freedman, S.S. Gubser, K. Pilch and N.P. Warner,
``Continuous distributions of D3-branes and gauged supergravity'',
JHEP {\bf 0007} (2000) 038, 
[hep-th/9906194]

\bibitem{Brandhuber:1999}
 A.~Brandhuber and K.~Sfetsos,
``Wilson loops from multicentre and rotating branes, mass gaps and
phase structure in gauge theories,''
Adv.\ Theor.\ Math.\ Phys.\  {\bf 3} (1999) 851
[hep-th/9906201].

%\cite{Anselmi:2000fu}
\bibitem{Anselmi:2000fu}
D.~Anselmi, L.~Girardello, M.~Porrati and A.~Zaffaroni,
``A note on the holographic beta and c functions,''
Phys.\ Lett.\ B {\bf 481} (2000) 346
[hep-th/0002066].
%%CITATION = HEP-TH 0002066;%%


\bibitem{mueck}
W.~M\"{u}ck,
``Correlation functions in holographic renormalization group flows,''
Nucl.\ Phys.\ B {\bf 620}, 477 (2002)
[hep-th/0105270].
%%CITATION = HEP-TH 0105270;%%

\bibitem{Notes} O. DeWolfe and D.Z. Freedman,
``Notes on Fluctuations and Correlation Functions in Holographic
Renormalization Group Flows'',
[hep-th/0002226].

\bibitem{theisen}
G.~Arutyunov, S.~Frolov and S.~Theisen,
``A note on gravity-scalar fluctuations in holographic RG flow  geometries,''
Phys.\ Lett.\ B {\bf 484} (2000) 295
[hep-th/0003116].
%%CITATION = HEP-TH 0003116;%%

\bibitem{Anatomy}  M. Bianchi, O. DeWolfe, D.Z. Freedman
and K. Pilch, 
``Anatomy of two holographic renormalization group flows,''
JHEP {\bf 0101} (2001) 021
[hep-th/0009156].


\bibitem{Lust}
G.~L.~Cardoso, G.~Dall'Agata and D.~Lust, ``Curved BPS domain walls and RG flow
in five dimensions,'' JHEP {\bf 0203}, 044 (2002) [hep-th/0201270];
%%CITATION = HEP-TH 0201270;%%
G.~Lopes Cardoso, G.~Dall'Agata and D.~Lust, ``Curved BPS domain wall solutions
in five-dimensional  gauged supergravity,'' JHEP {\bf 0107}, 026 (2001)
[hep-th/0104156].
%%CITATION = HEP-TH 0104156;%%

\bibitem{Sabra}
A.~H.~Chamseddine and W.~A.~Sabra,
``Einstein brane-worlds in 5D gauged supergravity,''
Phys.\ Lett.\ B {\bf 517}, 184 (2001)
[Erratum-ibid.\ B {\bf 537}, 353 (2002)]
[hep-th/0106092];
%%CITATION = HEP-TH 0106092;%%
A.~H.~Chamseddine and W.~A.~Sabra,
``Curved domain walls of five dimensional gauged supergravity,''
Nucl.\ Phys.\ B {\bf 630}, 326 (2002)
[hep-th/0105207];
%%CITATION = HEP-TH 0105207;%%
S.~L.~Cacciatori, D.~Klemm and W.~A.~Sabra,
``Supersymmetric domain walls and strings in D = 5 gauged supergravity
coupled to vector multiplets,''
JHEP {\bf 0303}, 023 (2003)
[hep-th/0302218].
%%CITATION = HEP-TH 0302218;%%

\bibitem{Behrndt}
K.~Behrndt and M.~Cvetic,
``Bent BPS domain walls of D = 5 N = 2 gauged supergravity coupled to
hypermultiplets,''
Phys.\ Rev.\ D {\bf 65}, 126007 (2002)
[hep-th/0201272].
%%CITATION = HEP-TH 0201272;%%

\bibitem{Bak}
D.~Bak, M.~Gutperle, S.~Hirano and N.~Ohta,
``Dilatonic repulsons and confinement via the AdS/CFT correspondence,''
[hep-th/0403249].
%%CITATION = HEP-TH 0403249;%%

\bibitem{FG}
C. Fefferman and C. Robin Graham, ``Conformal Invariants'', in 
{\it Elie Cartan et les Math\'ematiques d'aujourd'hui} (Ast\'erisque, 1985) 
95.

\bibitem{A&S}
M.~Abramowitz and I.~A.~Stegun (Eds.), Handbook of Mathematical Functions, 
9th printing (Dover, New York, 1972). 

%\cite{McAvity:1995zd}
\bibitem{McAvity&Osborn}
D.~M.~McAvity and H.~Osborn,
%``Conformal field theories near a boundary in general dimensions,''
Nucl.\ Phys.\ B {\bf 455}, 522 (1995)
[cond-mat/9505127].
%%CITATION = COND-MAT 9505127;%%

\bibitem{Freedman}
A.~B.~Clark, D.~Z.~Freedman, A.~Karch and M.~Schnabl,
``The dual of Janus $((<:)\leftrightarrow(:>))$ an interface CFT,''
[hep-th/0407073].
%%CITATION = HEP-TH 0407073;%%


\bibitem{3pt}
M.~Bianchi and A.~Marchetti,
``Holographic three-point functions: One step beyond the tradition,''
[hep-th/0302019];
%%CITATION = HEP-TH 0302019;%%
M.~Bianchi, M.~Prisco and W.~Muck,
``New results on holographic three-point functions,''
JHEP {\bf 0311} (2003) 052
[hep-th/0310129];
%%CITATION = HEP-TH 0310129;%%
W.~Muck and M.~Prisco,
%``Glueball scattering amplitudes from holography,''
[hep-th/0402068].
%%CITATION = HEP-TH 0402068;%%

%\cite{Klebanov:2000hb}
\bibitem{Klebanov:2000hb}
I.~R.~Klebanov and M.~J.~Strassler,
``Supergravity and a confining gauge theory: Duality cascades and  
$\chi$SB-resolution of naked singularities,''
JHEP {\bf 0008} (2000) 052
[hep-th/0007191].
%%CITATION = HEP-TH 0007191;%%

\bibitem{Skenderis:2000in}
K.~Skenderis,
``Asymptotically anti-de Sitter spacetimes and their stress energy  tensor,''
Int.\ J.\ Mod.\ Phys.\ A {\bf 16}, 740 (2001)
[arXiv:hep-th/0010138].
%%CITATION = HEP-TH 0010138;%%










\end{thebibliography}
\end{document}